\documentclass{aa}

\usepackage{graphicx}
\usepackage{txfonts}
\usepackage{multirow}
\usepackage[]{hyperref}
\usepackage{natbib}
\usepackage[usenames,dvipsnames]{color}
\usepackage{xspace}
\usepackage[normalem]{ulem}
\usepackage{multicol}

\newcommand{\Msol}{\ensuremath{\,\mathrm{M_\odot}}\xspace}
\newcommand{\Msolend}{\ensuremath{\,\mathrm{M_\odot}}}

\newcommand{\Lsol}{\ensuremath{\,\mathrm{L_\odot}}\xspace}

\newcommand{\kk}{\,kK\xspace}
\newcommand{\kyr}{\,kyr\xspace}
\newcommand{\Myr}{\,Myr\xspace}

\newcommand{\kms}{\,km\,s$^{-1}$\xspace}

\newcommand{\GG}[1]{}
\newcommand{\asc}{$\alpha_\mathrm{sc}$ }
\newcommand{\aov}{$\alpha_\mathrm{ov}$ }

 \def\mso{\,\mathrm{M}_\odot}

 \def\llso{\log\, L/{\rm L}_\odot \,}
 \def\simle{\mathrel{\hbox{\rlap{\hbox{\lower4pt\hbox{$\sim$}}}\hbox{$<$}}}}
 \def\simgr{\mathrel{\hbox{\rlap{\hbox{\lower4pt\hbox{$\sim$}}}\hbox{$>$}}}}

\begin{document}

   \title{Constraining mixing in massive stars in the Small Magellanic Cloud}

   \author{A. Schootemeijer 
            \and
            N. Langer
            \and 
            N. J. Grin
            \and
            C. Wang
            }
    \institute{Argelander-Institut f\"{u}r Astronomie, Universit\"{a}t Bonn, Auf dem H\"{u}gel 71, 53121 Bonn, Germany\\ \email{aschoot@astro.uni-bonn.de}
                \\
             }
             
    \authorrunning{Schootemeijer et al.}         

 
  \abstract
   {The evolution of massive stars is strongly influenced by internal mixing processes such as semiconvection, convective core overshooting, and rotationally induced mixing. None of these processes is currently well constrained.} 
   {We investigate models for massive stars in the Small Magellanic Cloud (SMC), for which stellar wind mass loss is less important than for their metal-rich counterparts. We aim to constrain the various mixing efficiencies by comparing model results to observations.}
   {For this purpose, we use the stellar evolution code MESA to compute more than 60 grids of detailed evolutionary models for stars with initial masses of
$9\dots 100\mso$, assuming different combinations of mixing efficiencies of the various processes in each grid. Our models evolve through core hydrogen and helium burning, such that they can be compared with the massive main sequence and supergiant population of the SMC. }
   {We find that for most of the combinations of the mixing efficiencies, models in a wide mass range 
spend core-helium burning either only as blue supergiants, or only as red supergiants. 
The latter case corresponds to models that maintain a shallow slope of the hydrogen/helium (H/He) 
gradient separating the core and the envelope of the models. Only a small part of the mixing parameter space 
leads to models that produce a significant number of blue and red supergiants, which both exist abundantly in the SMC. 
Some of our grids also predict a cut-off in the number of red supergiants above $\llso = 5$\dots$5.5$.
Interestingly, these models contain steep H/He gradients, as is required to understand the hot, 
hydrogen-rich Wolf-Rayet stars in the SMC. We find that unless it is very fast, 
rotation has a limited effect on the H/He profiles in our models.}
{While we use specific implementations of the considered mixing processes, 
they comprehensively probe the two first order structural parameters, 
the core mass and the H/He gradient in the core-envelope interface. Our results imply that, in massive stars, mixing during the
main sequence evolution leads to a moderate increase in the helium core masses, and also that the H/He gradients above
the helium cores becomes very steep. Our model grids can be used to further refine 
the various mixing efficiencies with the help of future observational surveys of the massive stars in the SMC, 
and thereby help to considerably reduce the uncertainties in models of massive star evolution.}


   \maketitle
%

\section{Introduction \label{sec:introduction}}
Massive stars play a central role in astrophysics. They dominate the evolution of star forming galaxies by providing chemical enrichment, ionizing radiation and mechanical feedback \citep[e.g.,][]{Hopkins14}. Massive stars also produce a variety of transient phenomena, like supernovae (SNe), long-duration gamma-ray bursts (lGRBs) and merging black holes emitting gravitational waves. 
Excitingly, these events can be so bright that they are observable up to high redshift, 
allowing us to study the early universe. 
In fact, lGRBs \citep{Graham17} superluminous SNe \citep{Chen17, Schulze18}, and likely also massive black hole mergers \citep{Belczynski10} appear predominantly in low-metallicity galaxies.
The implication is that massive star evolution can proceed differently in the early universe compared to
that in the Milky Way. However, at high redshift, massive stars can not be observed individually until they explode.

It is therefore important to study the existing low-metallicity massive stars that are nearby,
which are concentrated in star forming dwarf galaxies. A unique environment for this is provided
by the Small Magellanic Cloud (SMC), which, at a distance of $\sim$60\,kpc, allows the detailed
study of individual stars. With a metal content of about one fifth of the solar value \citep{Venn99, Korn00},
it is representative of massive star forming galaxies at a redshift around $z=3$ \citep{Kewley07}.

Evolutionary models of massive stars are plagued by our ignorance of two physical ingredients, mass loss and internal mixing. 
Also in this respect, it is beneficial to focus on the low
metallicity environment of the SMC, where stellar winds appear to be significantly weaker than in the Milky Way \citep{Mokiem07}.
For example, current evolutionary models predict that stars above $\sim$25\Msol in
 the Galaxy lose more than 10\% of their initial
mass by stellar winds during the main sequence evolution, while in the SMC this happens only for stars above $\sim$60\Msol \citep{Brott11}. By considering massive stars in the SMC, we aim to exploit this feature, which allows us, at least below 
a certain threshold mass, to focus on internal mixing as the major uncertainty in massive single star evolution. 


The evolution of massive stars is known to sensitively depend on a number of internal mixing processes \citep{Langer12}. 
The most important one is certainly convection, and ``convective overshooting'', 
i.e., mixing at the boundaries of convective regions \citep{Maeder88, Alongi93}, which affects the core masses and lifetimes of all phases of massive star evolution. Furthermore, semiconvection is an important but not well-understood process that determines the timescale of mixing in layers with a stabilizing gradient in the mean molecular weight \citep{Langer83}. 
It regulates the hydrogen/helium (H/He) gradient at the core-envelope interface in massive stars, 
thereby sensitively influencing their post main sequence radius evolution \citep{Langer91, Stothers92, Langer95}.
It should be noted that a deep understanding of what drives a stellar model to a blue or red solution after core hydrogen burning is still absent. Various ideas have been proposed and discussed \citep[e.g.,][]{Hoppner73, Renzini92, Iben93, Sugimoto00, Stancliffe09} but a concensus is still not there.
Finally, rotationally induced mixing may affect the evolution of massive stars, 
at least for the fraction of them that rotates rapidly \citep{Maeder00, Heger00, Yoon06}.
These mixing processes will not only affect the evolution of the surface properties of massive stars, but also determine their internal structure and are therefore important, e.g., for realistic pre-SN models.


While the individual effects of the efficiency of these mixing processes on evolutionary models are well known, often since decades, it is highly uncertain which efficiencies are realistic.
At present, this
is difficult to gauge from first-principle multi-dimensional calculations \citep{Merryfield95, Grossman96, Canuto99, Zaussinger13}. 
At the same time,
even when the influence of mass loss on the evolution can be neglected, it is also difficult 
to constrain the mixing efficiencies observationally, as all three processes may act at the same time. 

Here, we take advantage of the increase in computing power from the last decades, which allows us to explore the whole realistic parameter space for 
for more than one mixing process
simultaneously.
We compute
66 grids of massive star evolutionary models with 11 initial masses per grid, i.e., a total of 726 evolutionary sequences, where we varied the assumptions on the efficiency of overshooting, semiconvection, and, by choosing different initial rotational velocities, of rotationally induced mixing. We then explored which combinations of mixing efficiencies lead to models that are consistent with the available observational constraints. 

A second recent, applicable development is a study of \cite{Schootemeijer18}, who investigated Wolf-Rayet (WR) stars in the SMC. These authors found that
the progenitor stars of apparently single WR stars contain an about ten times steeper H/He gradient
than the one emerging during core hydrogen burning 
as a result of the retreating convective core. They suggested that this steep gradient could be caused by internal mixing; thus, we explore in which part of the mixing parameter space we can reproduce this gradient.


In Sect.\,\ref{sec:method}, we explain the method that we used to obtain our results, which are presented in Sect.\,\ref{sec:results}. In Sect.\,\ref{sec:earlier_work} we compare our results with earlier work. We interpret the further implications of observed blue and red supergiant populations on the efficiency of overshooting and semiconvection in Sect.\,\ref{sec:obs}. In Sect.\,\ref{sec:validity} we discuss the robustness of our results and finally we present the conclusions of our work in Sect.\,\ref{sec:conclusions}.

\section{Method \label{sec:method}}
We used MESA\footnote{{\tt http://mesa.sourceforge.net/}} \citep[][version 10108]{Paxton11, Paxton13, Paxton15, Paxton18} to simulate our grid of stellar evolution models. MESA is a one-dimensional stellar evolution code that solves the stellar structure equations. For the physics assumptions we followed \cite{Schootemeijer18}, who in turn have adopted a.o. the SMC chemical composition and a wind mass loss recipe as in \cite{Brott11}. 
Below, we highlight the most important physics assumptions.

The choice of wind mass loss recipe depends on the temperature $T_\mathrm{eff}$ and surface hydrogen mass fraction $X_\mathrm{s}$. For hot stars ($T_\mathrm{eff} > 25$\kk) that are hydrogen rich ($X_\mathrm{s} > 0.7$) we adopted the prescription of \cite{Vink01}. For hot hydrogen-poor stars ($X_\mathrm{s} < 0.4$) we used the wind of \cite{Hamann95} divided by ten. We linearly interpolated between the predicted $\log \dot{M}$ given by both prescriptions in case $0.4 < X_\mathrm{s} < 0.7$. For cold stars ($T_\mathrm{eff} \lesssim 25$\kk) we used the prescription from \cite{Nieuwenhuijzen90} 
in case it predicts a mass loss rate higher than \cite{Vink01}. 
Due to its high opacity, iron is the main driver of stellar winds. 
We scaled all winds to the iron abundance rather than the metallicity $Z$. The stellar winds thus scale as $\dot{M} \propto (X_\mathrm{Fe} / X_{\mathrm{Fe,}\, \odot})^{0.85}$, where the factor 0.85 is the metallicity dependence found by \cite{Vink01}. Here, $X_{\mathrm{Fe,}\, \odot} = 0.00124$ \citep{Grevesse96}.

The initial composition of our models is based on various observations. The iron mass fraction $X_\mathrm{Fe, \, SMC}$ follows from \cite{Venn99}, who found that [Fe/H]$_\mathrm{SMC} = -0.4$. The mass fractions from the elements C, N, O, and Mg are those as listed in \cite{Brott11}. The helium mass fraction ($Y$) is 0.252. Finally, the hydrogen mass fraction is calculated as $X = 1 -Y - Z$.

We adopted the Ledoux criterion for convection. In regions where convective mixing occurs, we employed the standard mixing length theory \citep{Bohm58} with a mixing length parameter of $\alpha_\mathrm{MLT} = 1.5$. In superadiabatic regions that possess a stabilizing mean molecular weight ($\mu$) gradient, we assumed semiconvective mixing to occur. The efficiency of semiconvective mixing  in our models is controlled by the scaling factor $\alpha_\mathrm{sc}$ \citep{Langer83}. We explored the range $\alpha_\mathrm{sc} = 0.01, \ldots, 300$.
The mixing region above hydrogen-burning convective cores is extended by a distance of $\alpha_\mathrm{ov} H_{\rm P}$, where $H_{\rm P}$ is the pressure scale height at the convective core boundary (i.e., we use step overshooting).
In our models, we considered the range $\alpha_\mathrm{ov} = 0.0, \ldots, 0.55$. The range of initial masses that we explored is $M = 9, \ldots, 100$\Msolend.


   \begin{figure}
   \centering
   \includegraphics[width = \linewidth]{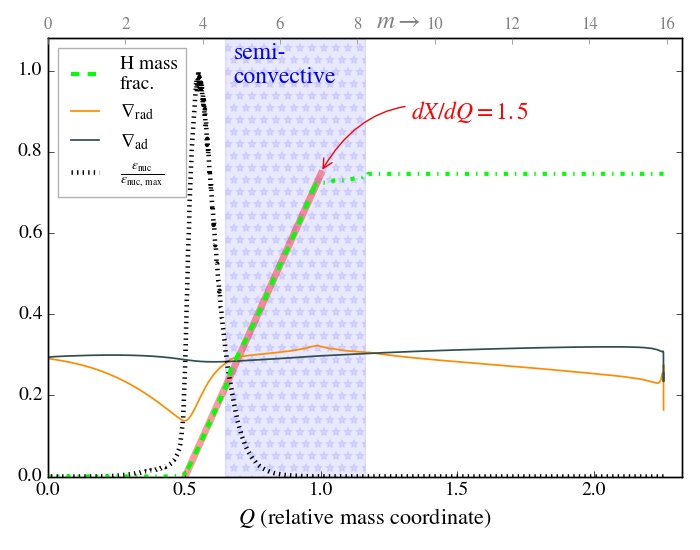}
   \caption{Various quantities of our 16\Msol stellar model computed with $\alpha_\mathrm{ov} = 0.11$ after core hydrogen depletion
as function of the internal mass coordinate. The model is undergoing hydrogen shell burning, as indicated by the relative rate of 
nuclear energy production (black dotted line). The hydrogen mass fraction is shown in green. 
The part of the hydrogen profile that was used to fit the H/He gradient $dX/dQ$ (see main text) is shown with a dashed line, 
the rest with a dot-dashed line. 
The resulting fit is shown as a straight red line. The semiconvective region, where 
   the radiative temperature gradient $\nabla_\mathrm{rad}$ exceeds the adiabatic temperature gradient $\nabla_\mathrm{ad}$
   in the presence of a stabilizing mean molecular weight gradient, is shaded in blue.}
             \label{fig:profile}%
    \end{figure}

We investigated which spatial and temporal resolution gives the most convergent results. We found that high spatial resolution ({\tt mesh\_delta\_coeff = 0.3} in MESA lingo 
-- this is a factor used to obtain the maximum difference in the log of the pressure, temperature, and helium mass fraction between two adjacent cells; it is multiplied by the default values of \cite{Paxton11})
and moderate time resolution ({\tt varcontrol\_target = 7d-4}) accomplishes this, i.e., our results appear unaffected by the resolution in the non-rotating case. 
It is known that different numerical choices, for a.o. time resolution, affect the exact amount of rotational mixing 
that takes place \citep{Lau14}. Therefore, we chose the MESA standard value for time resolution for our models 
including rotation ({\tt varcontrol\_target = 1d-4}.)


   \begin{figure*}
   \centering
   \includegraphics[width = 0.94\linewidth]{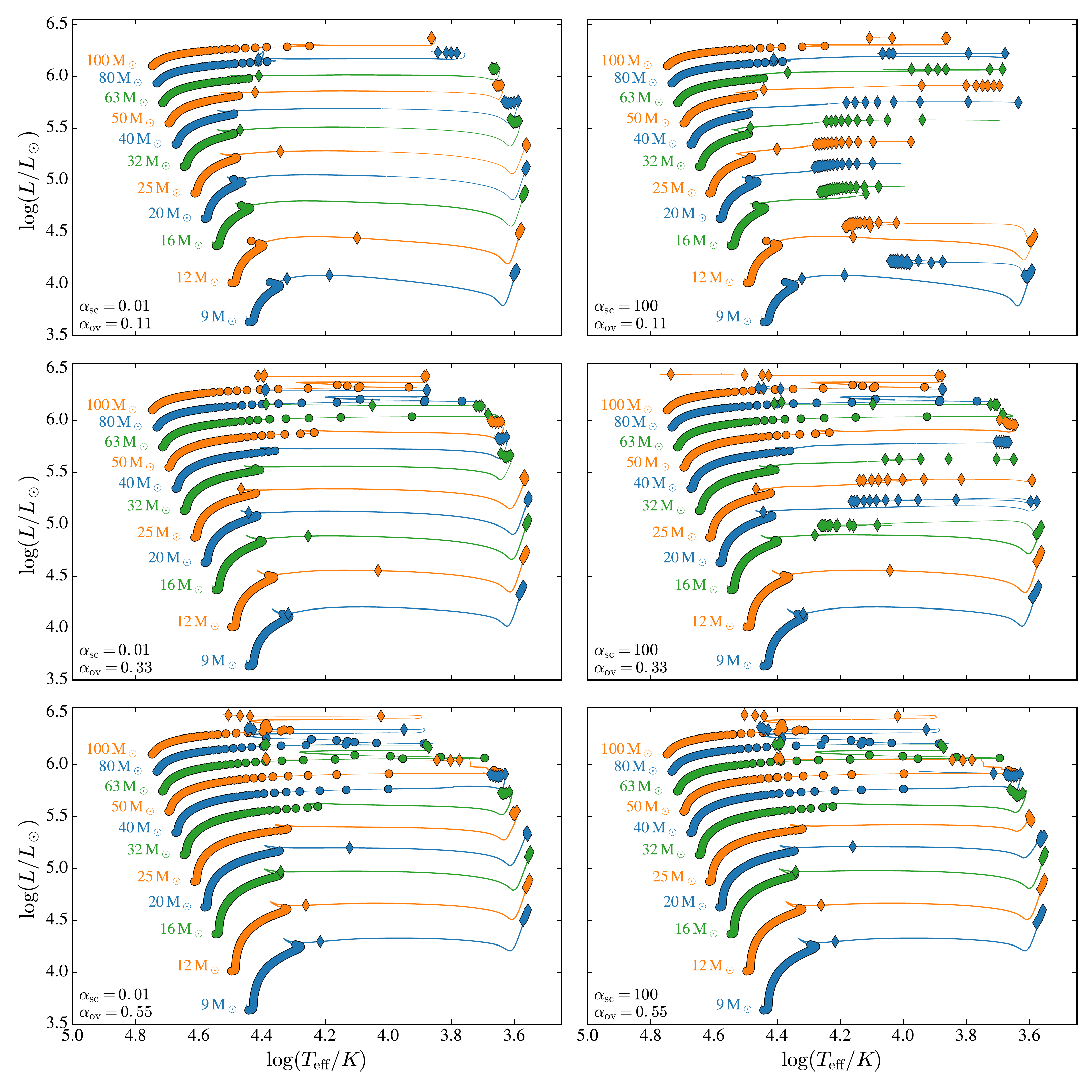}
   \caption{Evolutionary tracks in the Hertzsprung-Russell diagram for models computed with different efficiencies of overshooting and semiconvective mixing, from core hydrogen ignition to core helium exhaustion. The left panel shows models computed with inefficient semiconvective mixing ($\alpha_\mathrm{sc} = 0.01$) for three different overshooting parameters ($\alpha_\mathrm{ov} =$ 0.11, 0.33, and 0.55), while the right panels show models computed with efficient semiconvection ($\alpha_\mathrm{sc} = 100$), for the same three different values of the overshooting parameter. The time difference between two neighboring markers on a track is $50\,000\,$yr. A circular marker means that the model is core hydrogen burning, a diamond means indicates core helium burning. 
   The shown models are non-rotating.
   } \label{fig:six_hrds}%
    \end{figure*}


   \begin{figure}
   \centering
   \includegraphics[width = \linewidth]{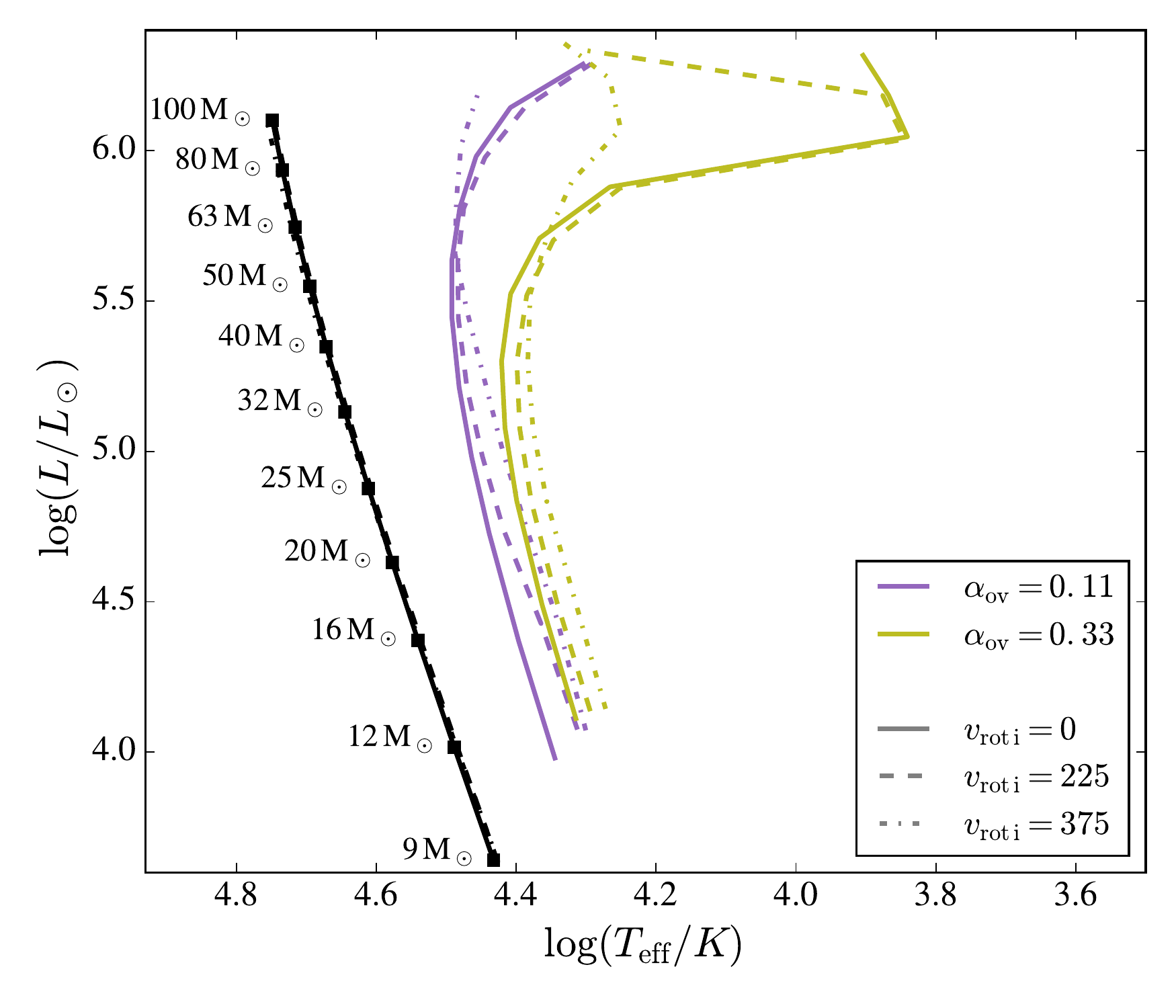}
   \caption{Hertzsprung-Russell diagram showing the terminal-age main sequence (TAMS) lines emerging from model grids computed with 
different combinations of overshooting \aov and initial rotation velocity $\varv_\mathrm{rot \, i}$ (in units of \kms). 
The semiconvective mixing efficiency is fixed to $\alpha_\mathrm{sc} = 1$, but the TAMS-lines are insensitive to 
this parameter. The black lines indicate the corresponding zero age main sequences.}
             \label{fig:tams_hrd}%
    \end{figure}

As an example of how we measure the slope of the H/He-gradient, 
we show a model from our grid in Fig.\,\ref{fig:profile} that has just exhausted hydrogen in its core. 
The retreat of the convective core during the main sequence evolution has left a nearly linear H/He gradient, 
which is well represented by a constant slope of $dX/dQ = 1.5$ (red line in Fig.\,\ref{fig:profile}). Here, $Q$ is a relative mass coordinate 
with $Q=1$ corresponding to the point where the linearly approximated hydrogen-profile reaches the initial hydrogen mass fraction 
of the star: $X = X_\mathrm{ini}$ \citep[see][]{Schootemeijer18}. 

Our evolutionary sequences are discontinued upon core helium exhaustion ($Y_\mathrm{c} = 0.01$). Stars in later, short-lived burning phases 
are not expected to constitute a significant fraction of massive star populations. 
While we do not compute binary models, we discuss the potential impact of this omission in Sect.\,\ref{sec:validity}.

\section{Results \label{sec:results}}

Below, we investigate in Sect.\,\ref{sec:hrds} the main effects that overshooting and semiconvection have on the evolution 
of massive star models in the Hertzsprung-Russell diagram (HRD). Subsequently, we discuss in more detail how overshooting, 
semiconvection and rotational mixing affect the internal chemical profiles of our models in Sect.\,\ref{sec:hhe_grad}.

\subsection{Effects of mixing on the evolution in the Hertzsprung- Russell diagram \label{sec:hrds}}

\subsubsection{Main sequence evolution}

The main mixing processes that play a role during the main sequence (MS) evolution are convection, overshooting, and rotational mixing. Semiconvection becomes important only after the main sequence stage (we elaborate on this in Sect.\,\ref{sec:hhe_grad}). 
As is well known \citep[e.g.,][]{Cloutman80}, mixing of layers above the convective core increases the MS lifetime and allows stars 
to end core hydrogen burning at lower surface temperatures and higher luminosities. In most evolutionary models in the literature, 
such mixing is assumed to be due to core overshooting, but rotational mixing can have a similar effect \citep[see e.g.,][]{Heger00}.
The effects of overshooting described above do emerge in Fig.\,\ref{fig:six_hrds}, where the MS broadens as \aov is increased (i.e., from the top to bottom panels). In the same figure, it can be seen that the MS evolution is unaffected by the efficiency of semiconvection. 

We further demonstrate the effects of overshooting and rotational mixing on the MS evolution in Fig.\,\ref{fig:tams_hrd}, 
where we display the terminal-age main sequence (TAMS) for evolutionary sequences with different initial rotation velocities (0, 225 and 375 \kms). 
We see that rotation mostly leads to slightly lower TAMS temperatures, like overshooting. There are two reasons for this. First, the central mixing region can be extended, although this effect is typically small in our models (cf. Sect.\,3.2.4). 
Second, because of rotation, the temperatures decrease as a result of 
the centrifugal acceleration reducing the effective gravity \citep[see e.g.,][their figs.\,3]{Chieffi13, Koehler15}.
In our model sequences, this effect is stronger 
at the TAMS than at the zero-age main sequence (ZAMS), because during the MS evolution the ratio of rotational to critical rotational velocity increases.
This effect is reduced for the highest considered masses, where stellar winds induce a spin-down of the models.
At the higher masses, rotational mixing is predicted to be the most efficient. If strong enough, the effects of rotation can push stars to the hot side of the HRD \citep[e.g.,][]{Maeder87, Yoon05, Chieffi13}. As a result, the TAMS in Fig.\,\ref{fig:tams_hrd} bends to higher temperatures for the most massive stars with a higher initial rotation velocity.


\subsubsection{Post main sequence evolution \label{sec:post_ms_hrd_evolution}}


In many of our model populations, the ratio of blue to red supergiant lifetime during core helium burning depends strongly on the efficiency of semiconvective mixing during the early stages of hydrogen shell burning \citep[as discussed by, e.g.,][]{Langer91, Stothers92, Langer95}. If such mixing is inefficient, the model sequences tend to favor 
red supergiant (RSG) solutions. This is illustrated in Fig.\,\ref{fig:six_hrds}, where the left panels show that all model sequences with $\alpha_\mathrm{sc} = 0.01$ (i.e., very inefficient semiconvection) spend virtually their entire helium core burning lifetime in a narrow effective temperature range as RSGs. In case semiconvection is very efficient ($\alpha_\mathrm{sc} = 100$, right panels in Fig.\,\ref{fig:six_hrds}), the lifetime in the blue supergiant (BSG) regime 
is enhanced. In particular when efficient semiconvection is combined with low overshooting, 
our evolutionary sequences spend most of core helium burning as BSGs. For large values for $\alpha_\mathrm{ov}$, 
the difference between low and and high \asc vanishes, because in that case semiconvective regions rarely develop in the deep hydrogen envelope. 
We discuss this in more detail in Sect.\,\ref{sec:overshooting}. 
Appendix\,A provides a figure similar to Fig.\,\ref{fig:six_hrds} that shows more combinations of \asc and $\alpha_\mathrm{ov}$ (Fig.\,\ref{fig:thirty_hrds}).


   \begin{figure}
   \centering
   \includegraphics[width = \linewidth]{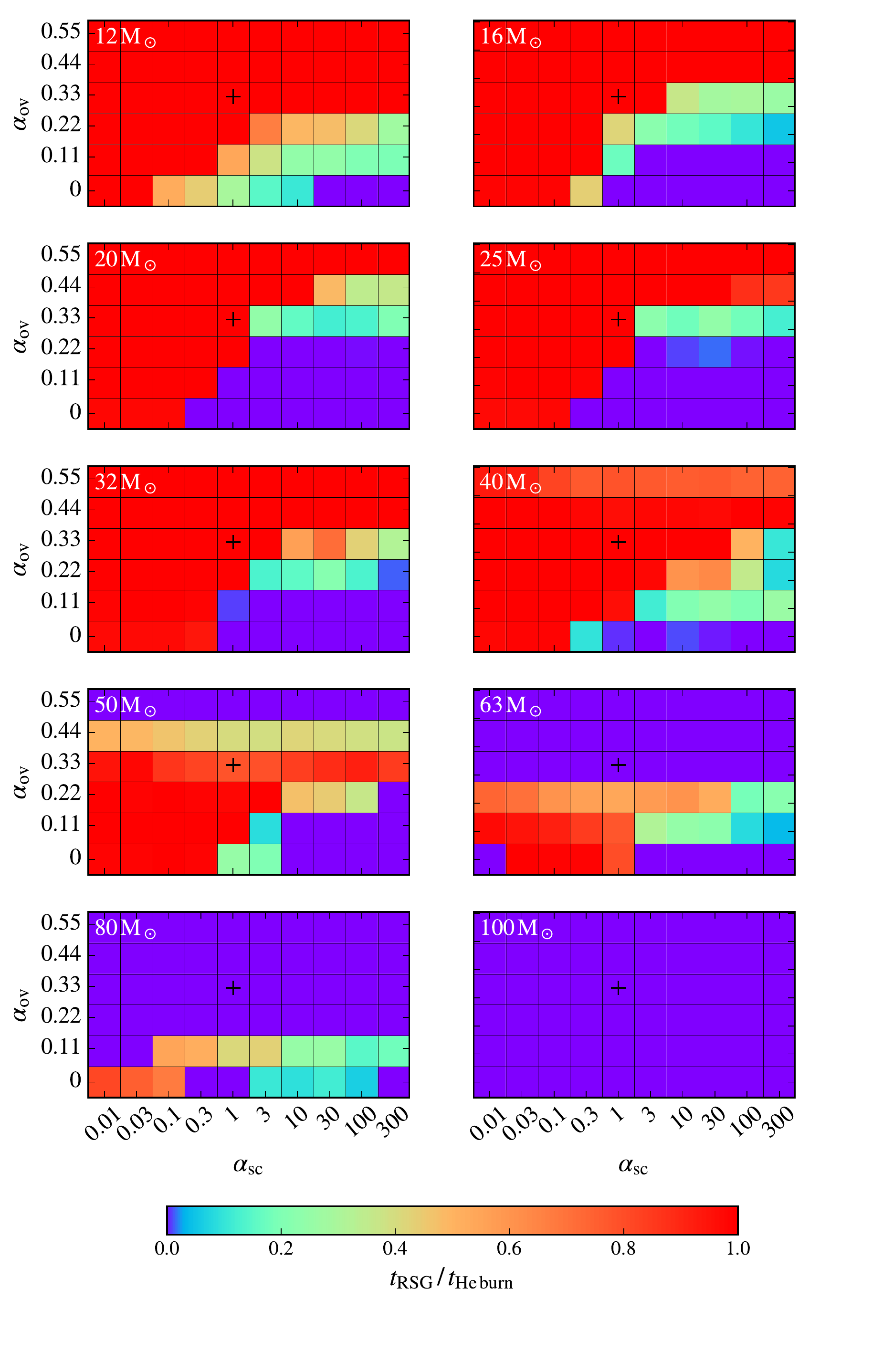}
   \caption{
   The fractional core helium burning lifetime that our model sequences spend as a red supergiant (color coded),
   as function of the adopted values of \asc and $\alpha_\mathrm{ov}$ in our model grids, for ten different
   considered initial masses. Each pixel represents one stellar evolution sequence. The cross indicates the
   parameters chosen in the models of \cite{Brott11}.
   }
    \label{fig:rsg_frac}%
    \end{figure}

To further quantify how our evolutionary models evolve during core helium burning, we compute the fraction 
$t_\mathrm{RSG}/t_\mathrm{He\,burn}$ of time spent as an RSG. 
Following \cite{Drout09}, we adopt a temperature threshold for RSGs as $T_\mathrm{eff} < 4800$\,K. 
The result is shown in Fig.\,\ref{fig:rsg_frac}.
We find that if semiconvection is inefficient ($\alpha_\mathrm{sc} \leq 0.3$) or overshooting is high ($\alpha_\mathrm{ov} \geq 0.44$), 
our models tend to be RSGs during core helium burning.
The only exception to this 
occurs at the highest masses: in that case, the more massive a star is and the larger the overshooting, the less time it spends as RSG. 
The reason is that the stellar winds in these models (which become stronger with mass) are able to remove a significant fraction of the hydrogen envelope (which becomes less massive with higher overshooting). If the mass of the hydrogen envelope becomes small enough, the models predict temperatures high enough for them to appear yellow or blue.


Models in the bottom right of the plots in Fig.\,\ref{fig:rsg_frac}, i.e., with low overshooting 
($\alpha_\mathrm{ov} \leq 0.22 $) and efficient semiconvection, often fail to reach RSG 
temperatures during core helium burning.
Models with efficient semiconvection and intermediate overshooting (around $\alpha_\mathrm{ov} = 0.22$ or 0.33, depending on mass) 
spend helium burning partly as RSG and partly as BSG. This can happen in two different ways. The first possibility, 
which we see in models with initial masses up to $\sim$20\Msol, is that after becoming RSGs they experience a blue loop excursion. 
The second possibility is that stars remain blue after core hydrogen exhaustion, and only become cooler later on --- 
as seen in some models with initial masses of $\sim$25\Msol or more. Both behaviors are present in Fig.\,\ref{fig:six_hrds}. 

Figure\,\ref{fig:six_hrds_rot} in Appendix\,A compares the evolutionary tracks from selected model grids for models with
initial rotational velocities of 225\,km/s and 375\,km/s. While minute differences can be seen, we find that 
rotation has no significant effect on the tracks for the majority of our models. We discuss this result further in Sect.\,\ref{sec:rotmix}. 


\subsection{The hydrogen/helium gradient \label{sec:hhe_grad}}
%


   \begin{figure}
   \centering
   \includegraphics[width = \linewidth]{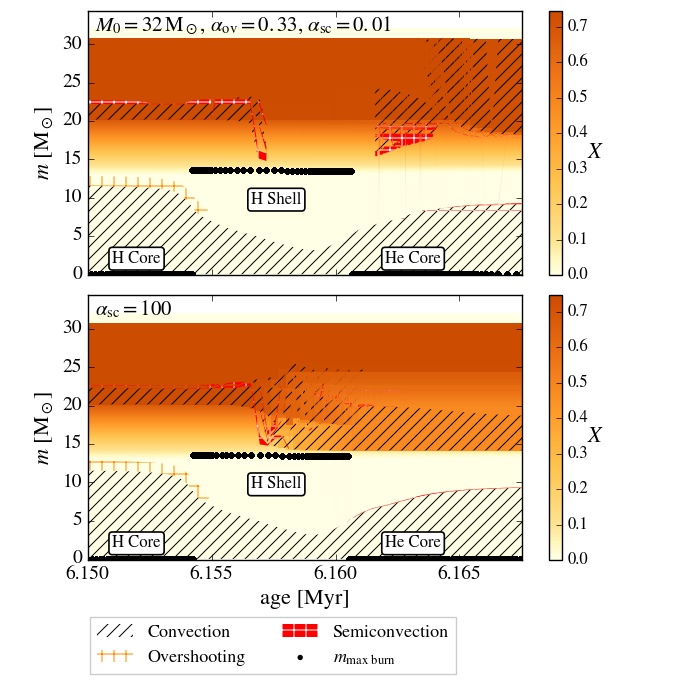}
   \caption{Kippenhahn diagram showing at what mass coordinates $m$ (y-axis) internal mixing regions arise, as well as the hydrogen mass fraction (color coded), as function of time, for two $32\mso$ evolutionary sequences. One is computed with
   inefficient semiconvective mixing ($\alpha_\mathrm{sc} = 0.01$; top panel), the other
   one computed with efficient semiconvective mixing ($\alpha_\mathrm{sc} = 100$; bottom panel).
   The displayed time interval starts near core hydrogen exhaustion and ends in the early stage of
   core helium burning.
   The overshooting parameter for both models is $\alpha_\mathrm{ov} = 0.33$,
   and rotation is not included.
   Black dots indicate the mass coordinate of the maximum specific nuclear energy generation.
   }

             \label{fig:kip}%
    \end{figure}

\subsubsection{Semiconvective mixing \label{sec:semiconvection}}
After core hydrogen exhaustion, massive star models undergo an overall contraction phase which leads to the ignition of hydrogen in a shell
. After this event, which for the two model sequences shown in Fig.\,\ref{fig:kip} occurs just before 6.155\Myr,
semiconvective regions form in the deep hydrogen envelope. The choice of \asc determines the efficiency of this 
mixing process in our models. 
When semiconvective mixing is inefficient (top panel of Fig.\,\ref{fig:kip}, where $\alpha_\mathrm{sc} = 0.01$) there is no 
significant composition change in the semiconvective regions above the hydrogen burning shell. As a result, 
the H/He gradient in the deep hydrogen envelope (e.g., as shown in Fig.\,\ref{fig:profile}) remains almost unaltered during
this phase.

On the other hand, when semiconvective mixing is efficient (bottom panel of Fig.\,\ref{fig:kip}, $\alpha_\mathrm{sc} = 100$) 
H/He gradients around mass coordinate $m = 15\dots20$\Msol can quickly wash out. As a result, the criterion for semiconvection
\begin{equation}
    \nabla_\mathrm{ad} < \nabla_\mathrm{rad} < \nabla_\mathrm{ad} + f \, \nabla_\mathrm{\mu} 
    \label{eq:criterion} 
\end{equation}
is no longer fulfilled.
In the equation above, $\nabla_\mathrm{ad}$ is the adiabatic temperature gradient ($d \log T / d \log P$)$_\mathrm{ad}$, $\nabla_\mathrm{rad}$ 
is the radiative temperature gradient and $\nabla_\mathrm{\mu}$ is the mean molecular weight gradient. 
Here, $f$ is defined as $f= - \chi_\mu / \chi_T$, where $\chi_\mu = (d\log P / d\log \mu)_{\rho, \, T}$ and $\chi_T = (d\log P / d\log T)_{\rho, \, \mu}$ .
Convection occurs in these layers after $\nabla_\mu$ has vanished. As a consequence of rapid semiconvective mixing, 
hydrogen-rich material is pushed close to the hydrogen-depleted core, thereby steepening the H/He gradient $dX/dQ$.

\begin{figure}
   \centering
   \includegraphics[width = \linewidth]{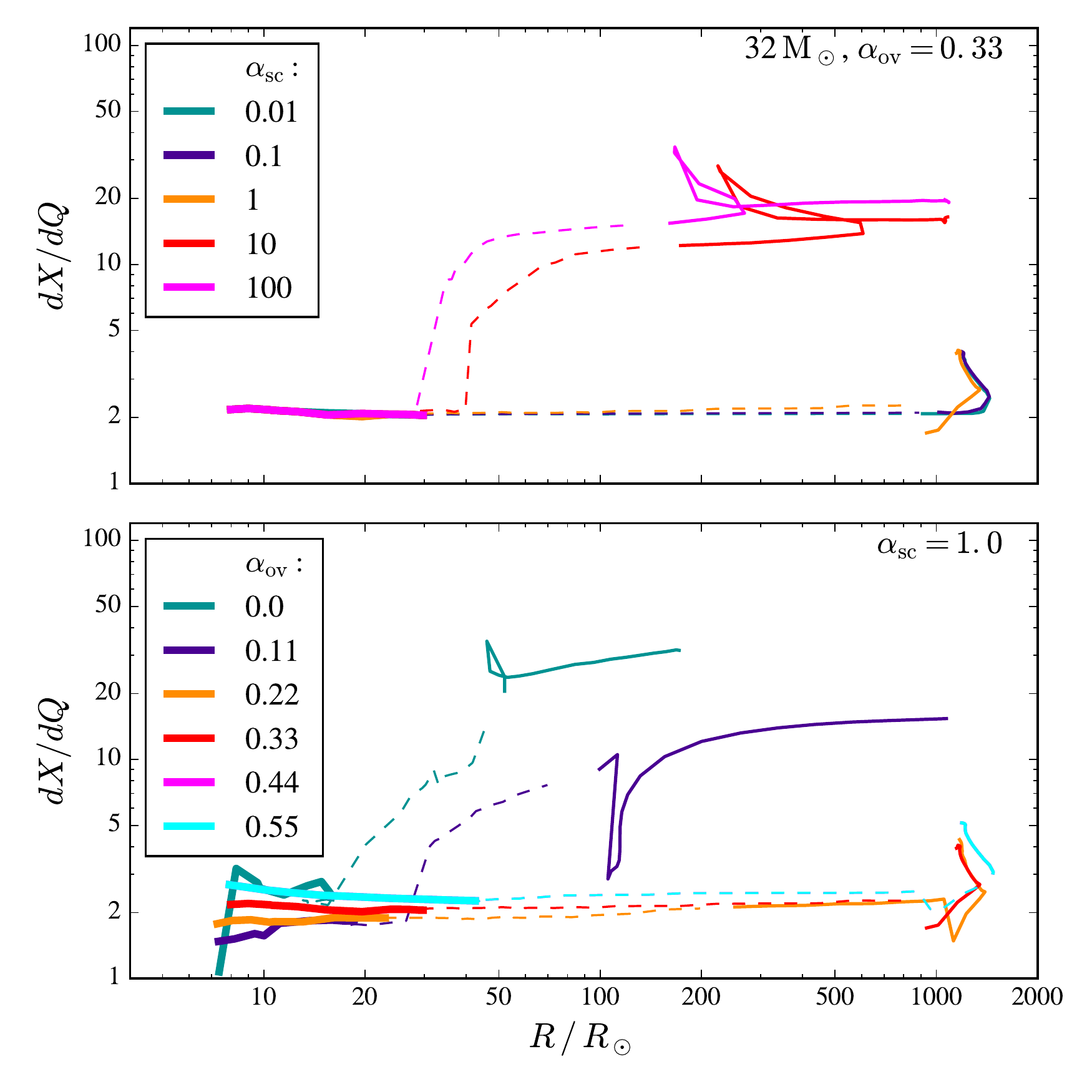}
   \caption{H/He gradient $dX/dQ$ (cf. Fig.\,\ref{fig:profile} in Sect.\,\ref{sec:method}) as a function of stellar radius
for 32\Msol sequences computed with various assumptions on internal mixing.
 Thick solid lines indicate core hydrogen burning, and thin solid lines indicate core helium burning. The short-lived in-between phase 
is displayed with a dashed line. In the top panel, $\alpha_\mathrm{ov} = 0.33$ is adopted in all models while \asc is varied. 
In the bottom panel, all models are computed with $\alpha_\mathrm{sc} = 1$ while \aov is varied.}
             \label{fig:logr_dxdq}%
    \end{figure}

The top panel of Fig.\,\ref{fig:logr_dxdq} shows $dX/dQ$ as a function of stellar radius $R$ for models computed with various semiconvective efficiencies and a fixed overshooting parameter ($\alpha_\mathrm{ov} = 0.33$).
In sequences with the most efficient semiconvection ($\alpha_\mathrm{sc} = 10, 100$
) the H/He gradient starts to increase immediately when the star expands after the MS. In contrast, the model sequences with less efficient semiconvection ($\alpha_\mathrm{sc} = 0.01, \, 0.1, \, 1$), experience no noticeable change in their hydrogen profile right after the MS. During core helium burning the H/He gradient only increases slightly as the innermost hydrogen layers are converted into helium.


\subsubsection{The role of overshooting \label{sec:overshooting}}


Similar to what we did in Sect.\,\ref{sec:semiconvection}, we explore here how the efficiency of overshooting affects 
the evolution of the radius $R$ and the H/He gradient $dX/dQ$. For this, we fix the semiconvection parameter to $\alpha_\mathrm{sc} = 1$. 
The bottom panel of Fig.\,\ref{fig:logr_dxdq} shows two effects of an increasing \aov on MS stars. 
First, $dX/dQ$ becomes slightly larger ($\sim$3 instead of $\sim$2), but not close to the $dX/dQ$ values of $\sim$10 or higher that are required to explain the properties of the apparently single WR stars in the SMC \citep[according to][]{Schootemeijer18}. Second, the stars can reach larger radii at the end of the MS. 
In the model sequence without overshooting, some semiconvective mixing can already occur during the MS, which causes a slight increase in $dX/dQ$. 

A third effect of overshooting manifests itself after the MS evolution. 
In model sequences with large \aov values, 
steep H/He gradients of $dX/dQ > 10$ do not develop. Overshooting plays a role here because it changes the shape of the hydrogen profile, 
which determines if and where the superadiabatic layers (i.e., where $\nabla_\mathrm{rad} > \nabla_\mathrm{ad}$, see Eq.\,\ref{eq:criterion}) form  that are required for semiconvective mixing. 
As found by \cite{Langer91}, such layers are less likely to form in models with larger overshooting, and as a result, less semiconvective mixing takes place in these models. 


   \begin{figure}
   \centering
   \includegraphics[width = \linewidth]{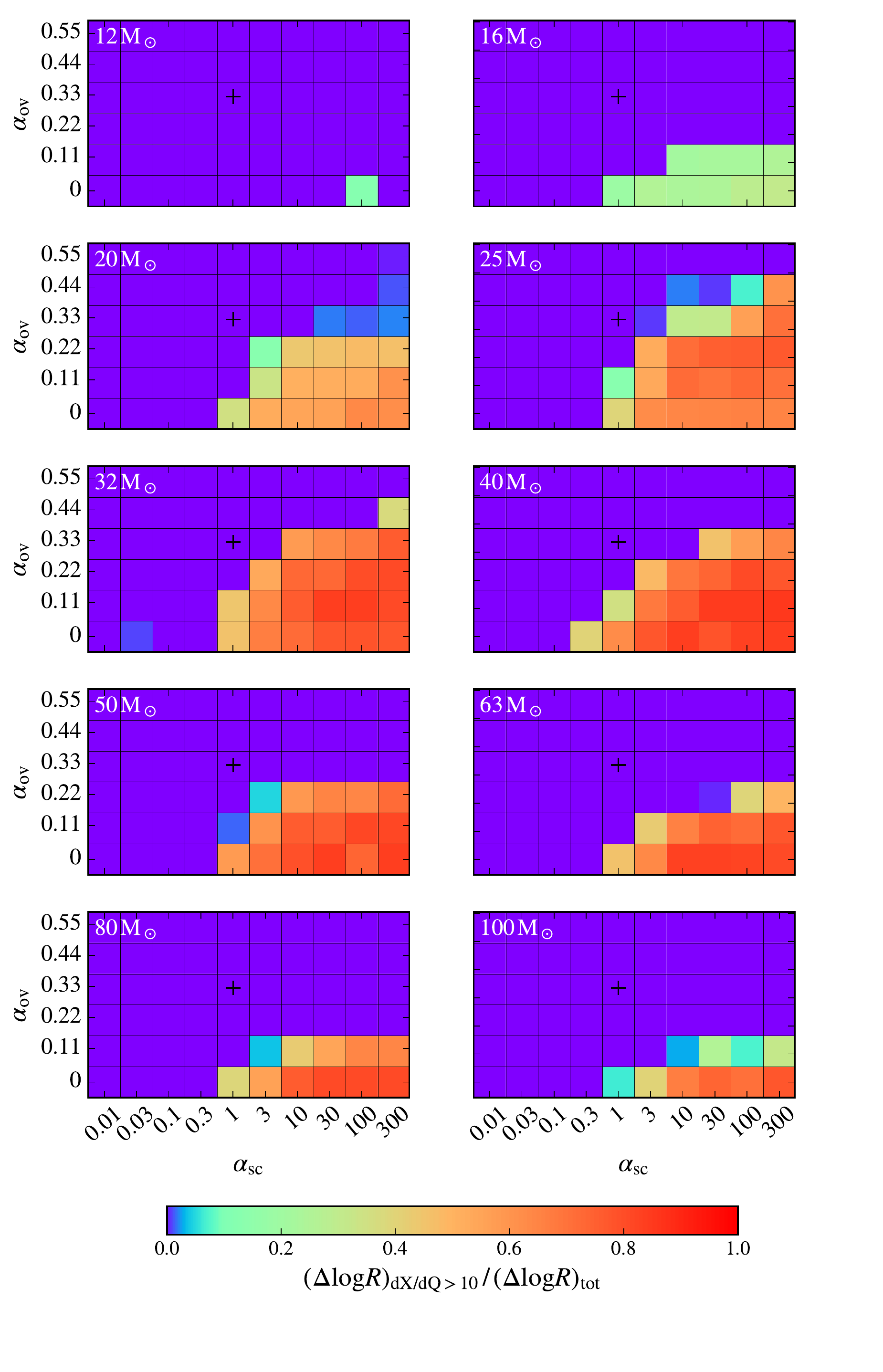}
   \caption{
   Same as Fig.\,\ref{fig:rsg_frac}, but now the color coding indicates during which fraction of the overall radius growth
   of a given sequence ($(\Delta \log R)_{\rm tot}$) the internal H/He gradient $dX/dQ$ exceeded a value of 10.   
   Blue color indicates models in which the H/He gradient remains shallow, while red color indicates models where
   a steep H/He gradient is established early on.
   }
             \label{fig:dxdq_sup10}%
    \end{figure}

This shows that overshooting has a strong effect on the amount of semiconvective mixing that takes place after core hydrogen exhaustion. 
Therefore, we consider the simultaneous variation of both mixing processes in the next section.

\subsubsection{Semiconvection and overshooting \label{sec:2d_results}}

Here, we consider the same model grids as displayed in Fig.\,\ref{fig:rsg_frac}.
Apart from the question \textit{if} models produce steep H/He gradients, we also want to answer the question \textit{when} 
they do so. This is especially important in the context of binary interaction.
Defining $(\Delta \log R)_\mathrm{tot}$ as the total increase in $\log R$ ($R$ is in Solar units here)
from the ZAMS to the maximum stellar radius, we consider which part of this increase occurs 
while the H/He gradient fulfills the criterion $dX/dQ > 10$:
$(\Delta \log R)_\mathrm{dX/dQ > 10}$.
For example, the $\alpha_\mathrm{sc} = 100$ model sequence shown in the top panel of 
Fig.\,\ref{fig:logr_dxdq} has a value of 
$(\Delta \log R)_\mathrm{dX/dQ > 10}\, / \, (\Delta \log (R)_\mathrm{tot} \approx 0.73$.

Fig.\,\ref{fig:dxdq_sup10} shows
that the majority of model sequences either never develop a steep H/He gradient
(blue pixels) -- this is the case for practically all models with $\alpha_\mathrm{sc} \leq 0.3$ 
as well as for almost all models with $\alpha_\mathrm{ov} \geq 0.44$ --
or they do so rather early during their post-MS expansion (orange/red pixels).
Only few sequences show intermediate behavior.


Only the models with the lowest shown initial masses (12\Msol and 16\Msolend) behave significantly different from what is described above. 
None of these are able to produce H/He gradients of $dX/dQ > 10$ before core helium ignition. In some model sequences (for $\alpha_\mathrm{sc} \geq 1$ and $\alpha_\mathrm{ov} \leq 0.11$)
such steep H/He gradients are reached during core helium burning, after the star has already expanded significantly -- therefore, they represent an intermediate case. 

The derived minimum initial mass of the hot SMC WR stars is $\sim$40\Msol \citep{Schootemeijer18}.
In this mass range, Fig.\,\ref{fig:dxdq_sup10} shows that to create steep H/He gradients inferred for SMC WR stars, the semiconvection efficiency needs to be of order unity or larger. In addition, overshooting can not exceed 0.33 pressure scale heights; in case their progenitors are born with 50\Msol or more, this number would be reduced to 0.22.

We find that the parameter space described above, where model sequences develop steep H/He gradients, 
is strongly correlated with the parameter space where model sequences spend at least a significant fraction 
of their helium burning lifetime as objects hotter than RSGs (Sect.\,\ref{sec:post_ms_hrd_evolution}). 
This shows that the occurrence of the post-MS BSG phenomenon is tightly linked to internal mixing.

\subsubsection{Rotational mixing \label{sec:rotmix}}

Rotation is predicted to drive mixing processes in the envelopes of rapidly spinning stars 
\citep[e.g.][]{Maeder87, Langer92}. As a result, these processes might have a non-negligible effect 
on the shape of the hydrogen profiles that we investigate. Therefore, we simulate a number of 
rotating model sequences to explore to what extent rotational mixing alters the hydrogen profile of our models.

\begin{figure}
   \centering
   \includegraphics[width = \linewidth]{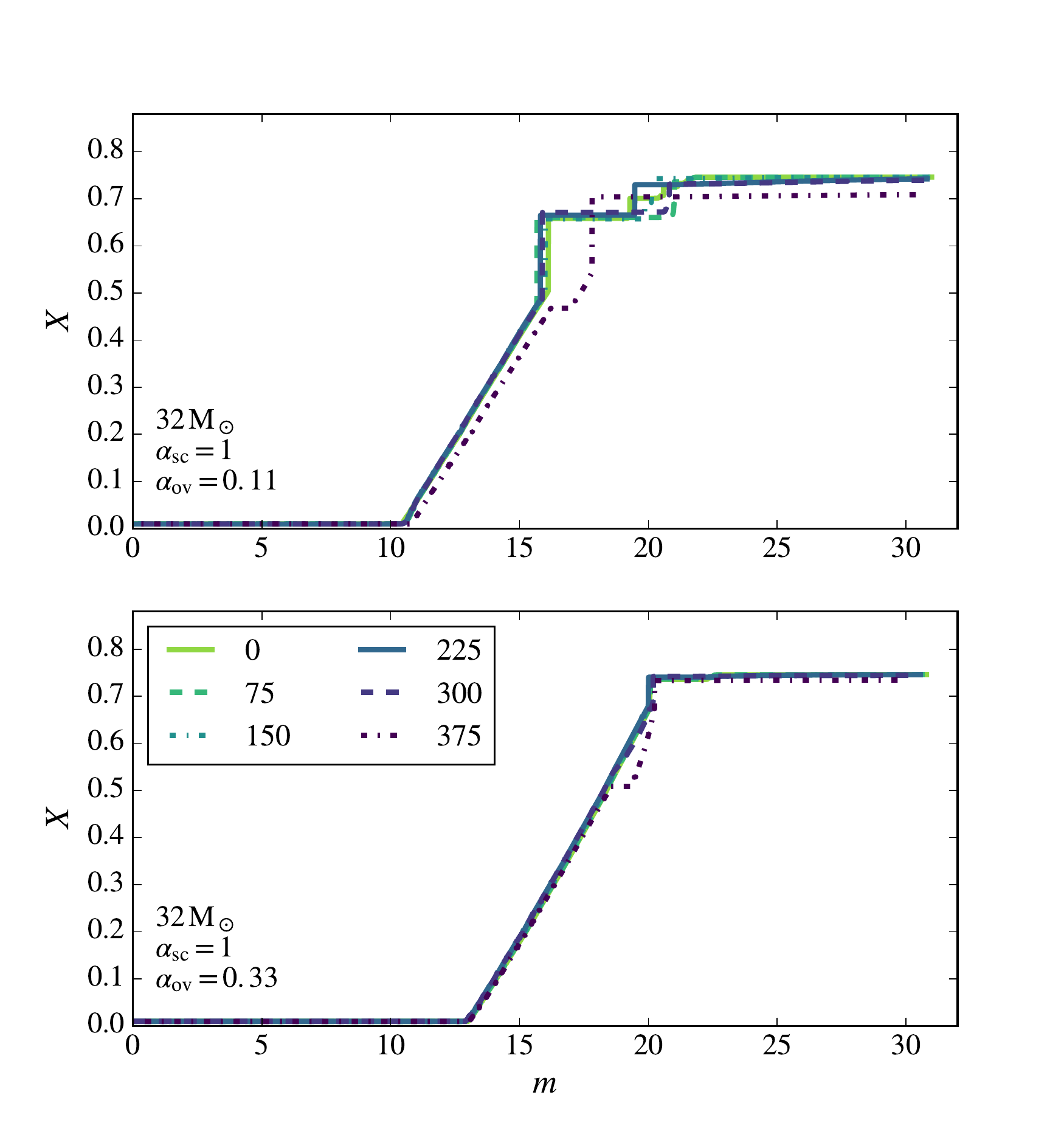}
   \caption{Hydrogen profiles of evolutionary models close to core hydrogen exhaustion, 
   for six different initial rotation velocities (given in \kms in the legend). The top panel shows evolutionary models 
   computed with an overshooting parameter of $\alpha_\mathrm{ov} = 0.11$, while  $\alpha_\mathrm{ov} = 0.33$ for the models shown in the bottom panel.
              \label{fig:vrot_profs}}%
    \end{figure}

In Fig.\,\ref{fig:vrot_profs}, we show the hydrogen profile of two sets of six models that are close to core hydrogen exhaustion ($X_\mathrm{c}=0.01$). These 32\Msol birth mass models have initial rotation velocities of $\varv_\mathrm{rot, \, i}  = 0, 75, \ldots \,, 375$\kms, 
and are computed with an overshooting parameter of either $\alpha_\mathrm{ov} = 0.11$ (top) or $\alpha_\mathrm{ov} = 0.33$ (bottom). 
The hydrogen profiles for the same value of overshooting are very similar. Only for the highest considered rotation velocity, 
$v_\mathrm{rot, \, i} = 375$\kms, a modest difference emerges.


O-type stars rotating with velocities of 375\kms appear to be rare in the SMC.
A study of 31 stars in the SMC cluster NGC 346 by \cite{Mokiem06} 
shows that less than one in five stars has a projected rotational velocity above $v \sin i > 200$\kms.
Furthermore, extreme rotators are predicted to evolve quasi-chemically homogeneously 
\citep[e.g.][]{Yoon05,Brott11}, in which case no sizable H/He gradient is expected. 
In conclusion, rotational mixing as implemented in our models does not appear to be a major factor in determining the shape 
of the chemical profile for the majority of stars and thus, it can not be expected be a major factor in both 
the formation of steep H/He gradients and in post-MS evolution. 
To illustrate the latter point, Fig.\,\ref{fig:six_hrds_rot} in Appendix\,A compares HRDs in which the evolutionary models have different initial rotation velocities.
However, we remind the reader that rotational mixing depends on a number of uncertain physics assumptions \citep{Lau14} and evolutionary models that include, e.g., stronger shear mixing, might show different behavior. We discuss these models in Sect.\,\ref{sec:earlier_work}.

\section{Comparison with earlier work \label{sec:earlier_work} }


Many of our physics assumptions are similar to the ones used by \cite{Brott11}, who adopted a semiconvection parameter of 
$\alpha_\mathrm{sc} = 1$, and who calibrated the overshooting parameter based on data from the FLAMES Survey of Massive Stars
\citep{Hunter08b} for stars of 16\Msol, for which they found $\alpha_\mathrm{ov} = 0.335$. 
Unsurprisingly, the post-MS evolution of these models and 
ours with the same \asc and $\alpha_\mathrm{ov} = 0.33$ is very similar -- both promptly ascend the red giant branch and do not experience blue loops. 
Also, in both sets of models, the smallest initial mass for which the TAMS bends to the coolest 
effective temperatures due to envelope inflation near the Eddington limit 
\citep{Sanyal15} is about 60\Msol.

Often in previous evolutionary calculations of massive star models, the Schwarzschild criterion has been adopted for convection. 
This assumption implies that stabilizing mean molecular weight gradients are not taken into account, which is equivalent to semiconvection leading to mixing as efficient as convection. 
Therefore, such evolutionary models might behave similar to our models for which we have 
used the largest semiconvective efficiency, i.e., $\alpha_\mathrm{sc} = 300$.
For example \cite{Charbonnel93, Meynet94} have published models using the Schwarzschild criterion, 
in combination with an overshooting parameter of $\alpha_\mathrm{ov} =0.2$.
\cite{Georgy13} produced similar models, but adopted 
$\alpha_\mathrm{ov} =0.1$. Below we compare the post-MS evolution of their low metallicity models 
with our results.

We find that the post-MS radius evolution of our $\alpha_\mathrm{sc} = 300$ (and 100), $\alpha_\mathrm{ov} = 0.22$ models is indeed similar to that of the $\alpha_\mathrm{ov} = 0.2$ models of \cite{Charbonnel93} and \cite{Meynet94}. 
In both cases, model sequences with initial masses up to $15-16$\Msol can become RSGs during core helium burning. This corresponds to an RSG upper luminosity of $\log(L / L_\odot) = 5.0$. More massive models, 
at least up to $30 - 40$\Msol, never reach RSG temperatures, or only in the very final moments of core helium burning.

\cite{Georgy13} provide models sequences ($\alpha_\mathrm{ov} = 0.1$) with and without rotation. Similar to what we describe above, the general behaviour of the non-rotating models is very similar to that of our $\alpha_\mathrm{sc} = 300$, $\alpha_\mathrm{ov} = 0.11$ model sequences.

A comparison of models including rotation is more difficult, as different implementations of the angular momentum and chemical transport processes are used in different groups.  
A treatment of rotation very similar to \cite{Heger05} and \cite{Suijs08} is implemented in the MESA code \citep{Paxton13}: angular momentum is transported by magnetic fields as suggested by \cite{Spruit02}. 
As a consequence, our models retain close to rigid rotation during their MS evolution,
making Eddington-Sweet circulations the dominant mixing process.  
As we have shown above, except for a slight reduction of the slope of the H/He gradient, the effects of rotational mixing remain rather limited
unless very fast rotation is considered.

\cite{Georgy13}, as well as \cite{Chieffi13} and \cite{Limongi18} neglect magnetic fields.
In their models, the shear instability dominates the transport of elements in radiative zones of the stellar models.
\cite{Martins13b, Choi16, Markova18, Limongi18} compare such models with those of \cite{Brott11} for Galactic metallicity, and find that 
more helium is mixed out of the core into the hydrogen-rich envelope in this case. As a consequence, larger helium cores are produced, similar to the effect of convective core overshooting, as well as shallower H/He gradients.
The finding of \cite{Georgy13} that their rotating and non-rotating models at $Z=0.002$ remain BSGs during essentially all of core helium burning (see their Fig.\,14) is consistent with 
the behaviour that is expected given their choice of $\alpha_\mathrm{ov} = 0.1$ and the Schwarzschild criterion for convection. The non-rotating models of \cite{Limongi18} produce mostly RSGs (for $[\mathrm{Fe/H}]\geq-1$, except at the highest masses), as they employ the Ledoux criterion and inefficient semiconvection.
With increasing rotation, the fraction of their models spent in the BSG regime appears to increase in the mass range
$15\dots 25$\Msol (cf., their Fig.\,14), in line with the additional rotational mixing in semiconvection zones mimicking a larger semiconvective mixing efficiency. 

While our analysis of the literature results must remain tentative, it does support the general idea
that the core mass and the H/He-gradient at the core-envelope interface are the dominant factors in determining the
post-MS radius evolution of massive star. This underlines the view that our results are meaningful
beyond the particular parametrization of the individual mixing process chosen for the models of our new stellar evolution
grids.



\section{Observational constraints \label{sec:obs}}

Above, we have described in which way the various considered internal mixing processes affect the observable stellar properties, i.e., the evolution of the models in the HRD and the H/He gradient. 
In this section, we attempt to find observational constraints which allow to rule out certain classes of models, and thereby constrain the efficiency of the internal mixing processes. 

\subsection{Main sequence stars\label{sec:ms_evolution}}

The occurrence of semiconvection during the MS evolution of massive stars has been
studied by, e.g., \cite{Chiosi70, Chiosi78, Langer85}, who showed that while semiconvection can occur prominently,
the evolution of the models during this stage is hardly affected by the choice of the semiconvective mixing efficiency.
Therefore, we concentrate here on the discussion of the effect of convective core overshooting, whose main effect is moving the location of the TAMS to cooler effective temperatures in the HRD \citep[e.g.][]{Prantzos86, Maeder88, Stothers92}.

To find valid constraints on convective overshooting for massive stars has proven difficult, since the isochrone method used for low and intermediate mass stars \citep{Maeder91} 
can not be used due to a lack of populous young star clusters. 
\cite{Brott11} used the distribution of rotational velocities to conclude that around 16$\Msol$, the overshooting
parameter should be close to $\alpha_\mathrm{ov} = 0.33$. 

By studying all the Galactic massive stars for which parameters have been determined through a detailed model
atmosphere analysis, \cite{Castro14} could compare their observable parameters with the predictions of stellar
evolution models. They confirmed a value of $\alpha_\mathrm{ov} = 0.33$ near 16$\Msol$ from their data,
but found that smaller values would be preferred for smaller initial masses, and larger ones for higher masses. 
Using their result, Grin et al. (in prep.) found that the best fit to the empirical TAMS of \cite{Castro14} is obtained 
for an increasing overshooting with mass such that $\alpha_\mathrm{ov} = 0.2$ at 8\Msol up to $\alpha_\mathrm{ov} = 0.5$ at 20\Msol, 
with a constant value for higher masses.

In a recent study, \cite{Castro18} analyzed the population of SMC OB field stars as observed within
the RIOTS4 survey \citep{Lamb16}. They could derive a tentative TAMS, which agrees well with the one from the \cite{Brott11} 
SMC models using $\alpha_\mathrm{ov} = 0.335$ while also finding some hints of a mass dependence.
 We conclude that, for the considered mass range, values of the overshooting parameter 
of the order of $\alpha_\mathrm{ov} = 0.3$ seem to provide the best match to observations of massive main sequence stars.


\subsection{Red supergiants \label{sec:rsg_observations}}

Both, \cite{Levesque06} and \cite{Davies13} show that the SMC RSG population extends up to a luminosity of $\log (L / L_\odot) \approx 5.3 - 5.4$,
but not beyond. Earlier studies \citep[e.g.,][]{Massey03b} 
reported a higher luminosity cut-off. However, the inferred temperatures have since been revised upwards, resulting in smaller bolometric corrections and hence lower luminosities.
The most recent investigation of RSGs in the SMC by \cite{Davies18}, who integrated photometric fluxes over a wide wavelength range to obtain the luminosity, also found a cutoff around $\log( L / L_\odot) \approx 5.4$. The latter study used a large sample of $\sim$150 RSGs with $\log( L / L_\odot) > 4.7)$. 

\begin{figure}
   \centering
   \includegraphics[width = \linewidth]{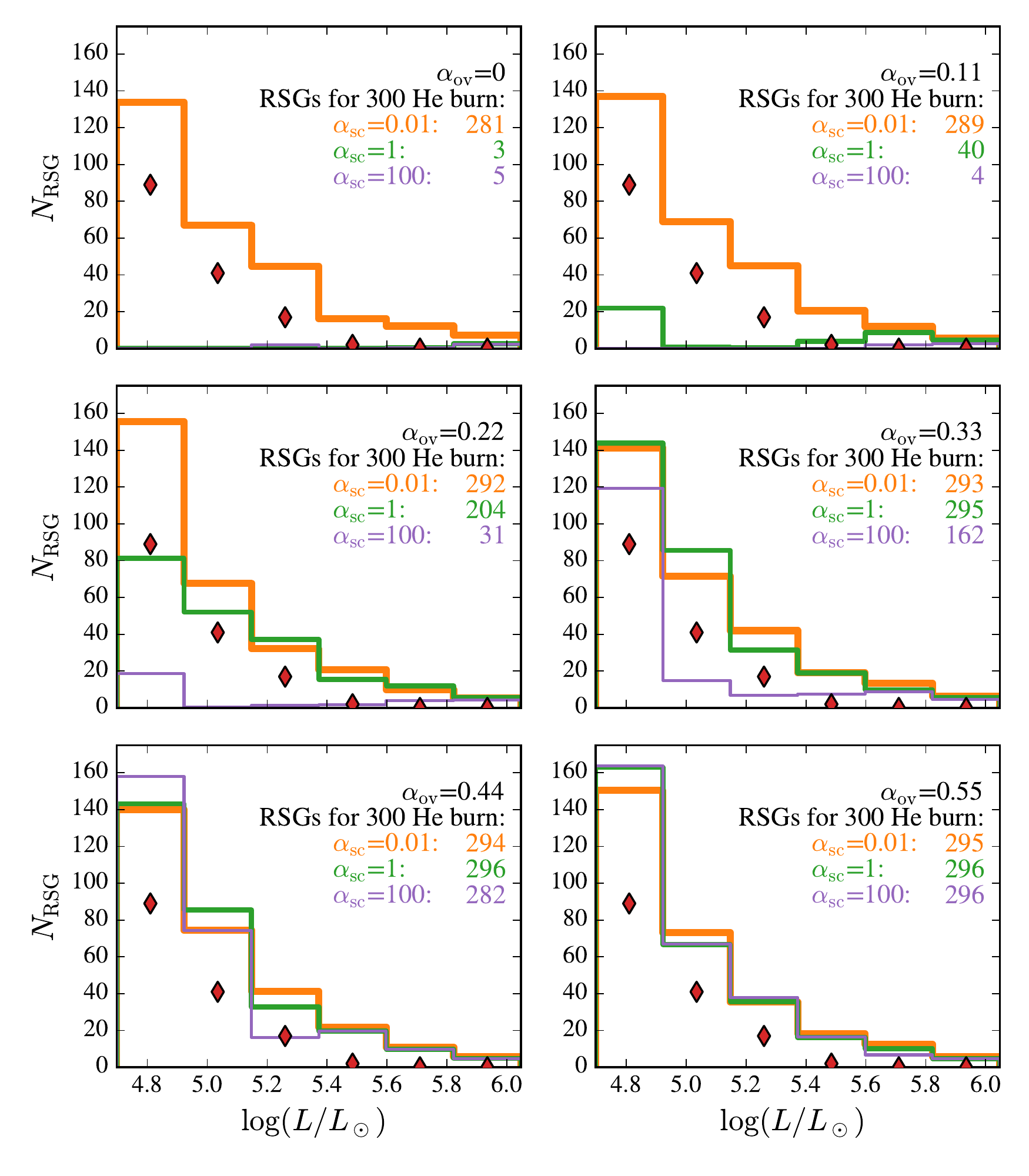}
   \caption{Diagrams showing the predicted luminosity distribution of red supergiants in the SMC. Each of the six diagrams has a different value for the overshooting parameter $\alpha_\mathrm{ov}$. The red diamonds indicate the number of red supergiants observed by \cite{Davies18}.}
             \label{fig:l_dist}%
    \end{figure}

In Fig.\,\ref{fig:l_dist}, we compare the RSG luminosity distribution of \cite{Davies18} to our theoretical predictions for different \asc and \aov combinations.
To obtain the theoretical predictions, we adopt a constant star formation rate and a Salpeter initial mass function \citep{Salpeter55}.
We further assume that there are 300 core helium burning stars in the SMC with $\log( L / L_\odot) > 4.7$. 
We obtain this number by adding the number of 
150 BSGs (see Sect.\,\ref{sec:bsg_observations})
in this luminosity range to the 150 RSGs in the sample of \cite{Davies18}.
Although uncertain, we expect this to match the true number of stars to within at least a factor of a few, 
as it could be a lower due to foreground contamination, or a higher due to to incompleteness of the sample, 
in particular for the BSGs. 
With this uncertainty in mind, the predicted distributions in Fig.\,\ref{fig:l_dist} could still be a good fit to the observations
even if they would need to be scaled up or down by some amount. To be consistent with \cite{Davies18}, we make the division between RSGs and hotter objects at 7\kk.


Figure\,\ref{fig:l_dist} shows that largest discrepancy between the observed RSG luminosity distribution and the predictions occurs for efficient semiconvection and the lowest overshooting, where hardly any RSGs are produced (see also Fig.\,\ref{fig:rsg_frac}). 
There, the RSG number fraction $ f_\mathrm{RSG} = N_\mathrm{RSG} / (N_\mathrm{RSG} + N_\mathrm{BSG})$ is only a few per cent, 
which would imply that about 1500-5000 BSGs would need to be present in the SMC given that 150 RSGs are observed. 
Such a large BSG population would rival the entire SMC in terms of stellar luminosity \citep[which is about $4\cdot10^8$\Lsol;][]{Bekki09}. \cite{Davies18} compare with tracks 
of \cite{Georgy13} where the Schwarzschild criterion 
for convection (which translates into extremely efficient semiconvection) and $\alpha_\mathrm{ov} = 0.1$ are assumed. 
Indeed, when we trace the progenitor evolution of the RSG models that were considered
we find that $\sim$5000 BSGs should be present to obtain 150 RSGs, based on the corresponding lifetimes of Georgy et al.'s models.


In model sequences where little semiconvective mixing occurs  (i.e., where \asc is low and/or \aov is high), 
the mismatch in Fig.\,\ref{fig:l_dist} between observations and model predictions is less striking. 
Up to $\log( L / L_\odot) = 5.3$, the observed steepness of the drop of the number of RSGs with increasing luminosity could reflect a combination of the initial mass function and the shorter lifetime of heavier helium cores. This drop can be well matched by the model sequences that burn helium as RSGs.
However, for luminosities of $\log( L / L_\odot) > 5.3$, essentially no RSGs are observed (Davies et al. list one RSG at $\log( L / L_\odot) \simeq 5.4$, and one more at $\log( L / L_\odot) \simeq 5.6$),
while grids of models where little semiconvective mixing occurs predict some tens of RSGs.

There is a limited range of mixing parameters that might be compatible with the observed paucity of luminous RSGs. We see in Fig.\,4 that in the mass range from 12 to 25$\Msol$, for efficient semiconvection, the pure BSG solutions tend to prevail more often when the mass increases.
Therefore, if nature had chosen for convective core masses corresponding to, e.g.,  $\alpha_\mathrm{ov}=0.22$ and $\alpha_\mathrm{sc}=100$, there would be a gap in the luminosity distribution of RSGs, because stars with initial masses of 20$\Msol$ and above would never become RSGs. 
Fig.\,\ref{fig:thirty_hrds} shows several such luminosity gaps for low overshooting and high semiconvection parameters, and Fig.\,\ref{fig:six_hrds_rot} shows that these gaps also prevail in our models which include rotational mixing. 
Such gaps could therefore explain the  observed scarcity of luminous RSGs with masses of the order of 25-32\Msol.
In this case the upper RSG luminosity limit 
would not be set by 
strong RSG winds stripping off the hydrogen envelopes in more luminous stars. Instead, it would be set by the helium burning stars around the upper luminosity limit being more likely to be BSGs.



   \begin{figure}
   \centering
   \includegraphics[width = \linewidth]{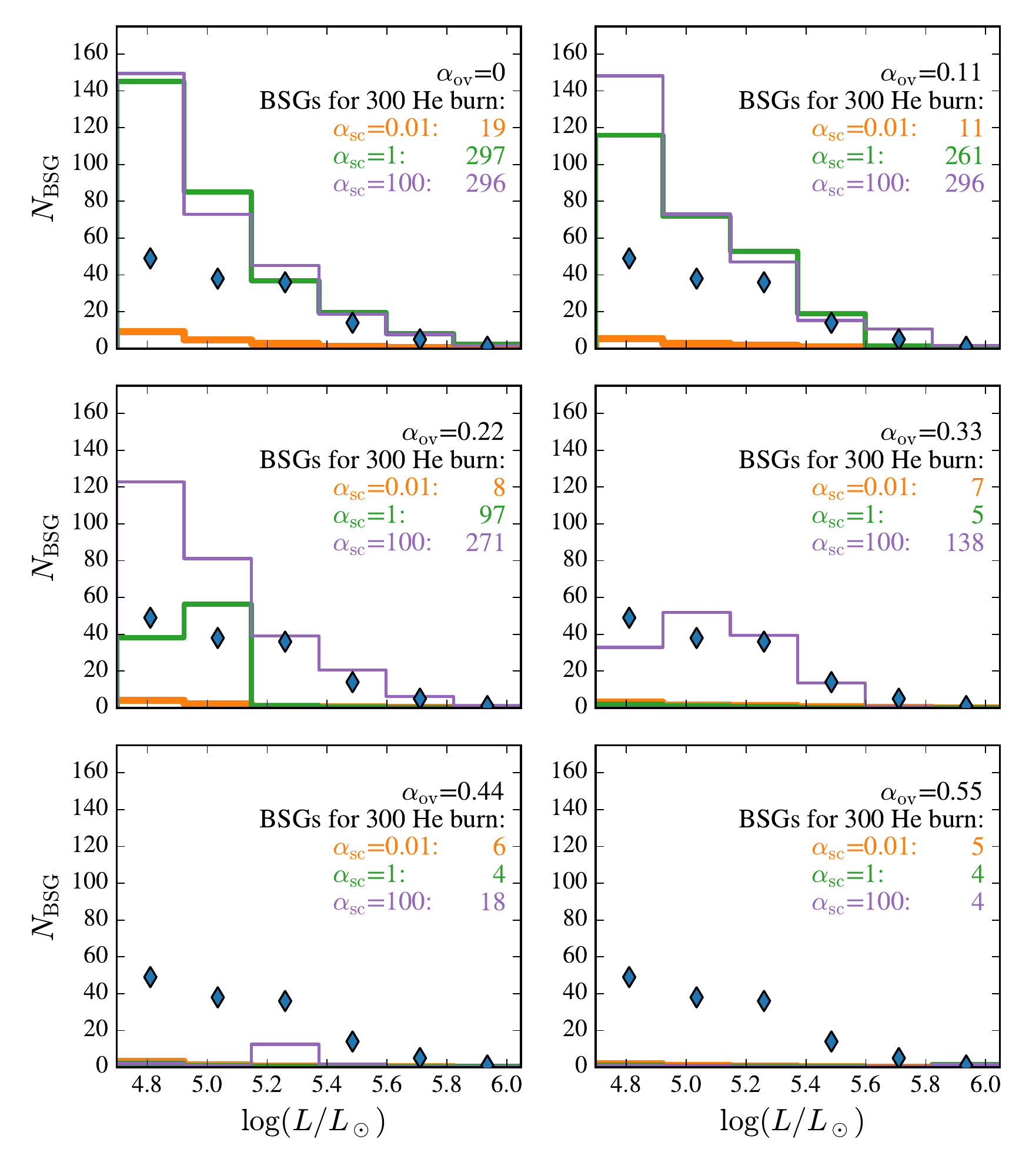}
   \caption{Same as Fig.\,\ref{fig:l_dist}, but this diagram is showing the luminosity distribution of blue supergiants. Here, blue diamonds indicate the combined number of blue supergiants observed in the SMC by \cite{Humphreys91} and \cite{Kalari18}. 
   }
             \label{fig:l_dist_bsg}%
    \end{figure}


In most of our grids, we find models above $\sim$40\Msol ($\log( L / L_\odot) \simeq 5.7$) to evolve to low effective temperatures, even though they remain hotter than classical RSGs. The fact that in these cases such objects are predicted (together, the two $\log( L / L_\odot) \geq 5.7$ bins in Fig.\,\ref{fig:l_dist} typically contain of the order of 15 to 20 objects) but zero are observed could have to do with uncertainties in mass loss rates 
for the brightest cool stars \citep[as discussed by][]{Davies18}. 
This may also be prone to the uncertain effect of envelope inflation \citep{Sanyal17}. 
Alternatively, stars this massive could simply happen to be rare in the SMC for a reason that is not yet fully understood \citep[as indicated by the results of][]{Blaha89}.

It remains remarkable that the range in mixing parameters which offers a tentative solution of the RSG luminosity problem described in \cite{Davies18} has a significant overlap with the parameter ranges favored by other observational constraints.



   \begin{figure*}
   \centering
   \includegraphics[width = \linewidth]{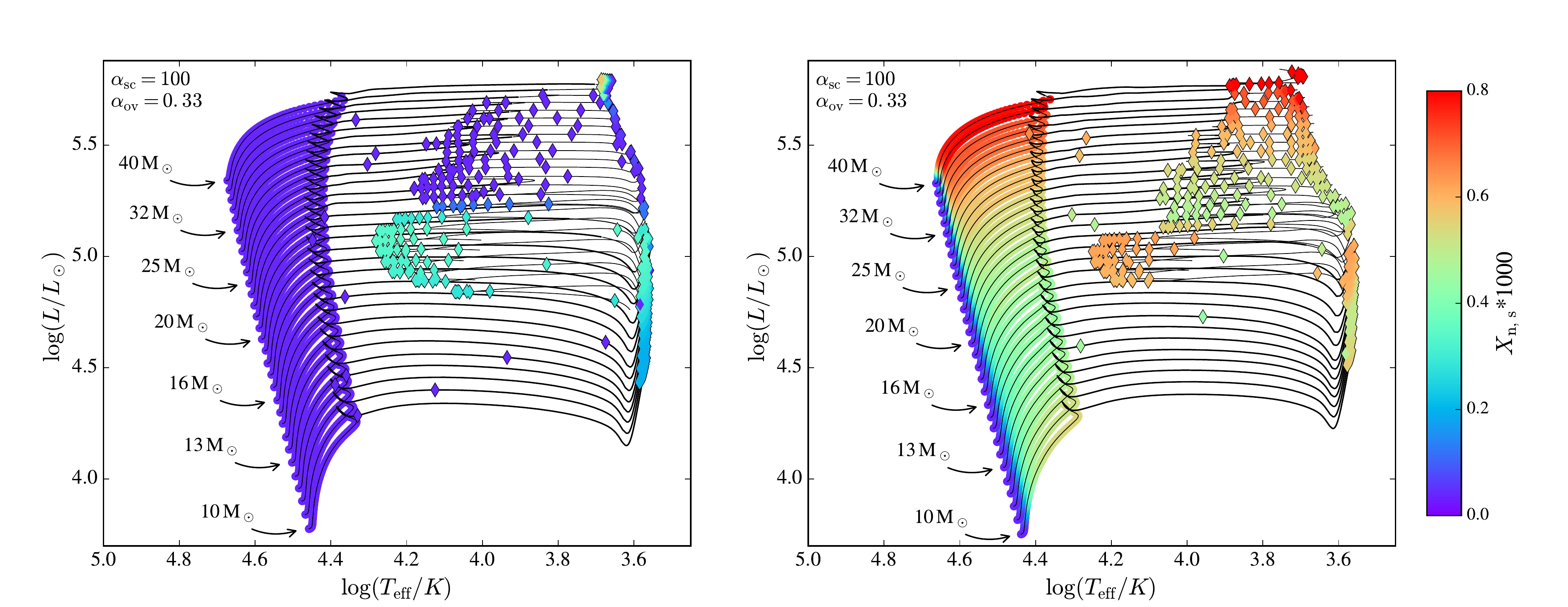}
   \caption{Hertzsprung-Russell Diagram showing the nitrogen surface mass fraction of our evolutionary sequences with $\alpha_\mathrm{sc} = 100$ and $\alpha_\mathrm{sc} = 0.33$. The initial masses of the shown stellar tracks are, logarithmically, evenly spaced every 0.02 dex. Similar to Fig.\ref{fig:six_hrds} a marker indicates a 50\,000\,yr timestep, a circle indicates core hydrogen burning and a diamand indicates core helium burning. \textbf{Left}: the model sequences are non-rotating. \textbf{Right}: the model sequences have an initial rotation velocity of 300\kms.}
             \label{fig:hrd_xns}%
    \end{figure*}

\subsection{Blue supergiants \label{sec:bsg_observations}}


The shape and size of the the BSG population in the SMC appears to be more uncertain than that of the RSG population. For our analysis, we combine the two samples of $\sim$100 A-type supergiants from \cite{Humphreys91} and $\sim$50 B-type supergiants of \cite{Kalari18}. 
This number is comparable to the number of RSGs; as a result, the blue-to-red supergiant ratio (b/r ratio) is then of order unity.
These samples cover the entire temperature range of BSGs, and we consider the same luminosity range as for RSGs (Sect.\,\ref{sec:rsg_observations}).
However, observational studies presented by \cite{Blaha89} and \cite{Evans06} also contain some tens of objects in the BSG temperature range -- thus, depending on the amount of overlap and other biases, the above number of BSGs could be seen as a lower limit.


Figure\,\ref{fig:l_dist_bsg} shows that in synthetic populations where efficient semiconvective mixing does not take place (e.g., if $\alpha_\mathrm{sc} = 0.01$ or $\alpha_\mathrm{ov} \geq 0.44$), the BSG population is negligible.
This means that the predicted b/r ratio is close to zero.
This seems to be at odds with aforementioned observations. For the opposite case, e.g., where $\alpha_\mathrm{sc} = 100$ and $\alpha_\mathrm{ov} < 0.22$, almost all helium burning stars are predicted to be BSGs, resulting in a b/r ratio much larger than unity.
While, given the uncertainties described above, this evolutionary scenario could still agree with the observed BSG population, it is clearly challenged by the presence of the observed RSG population (Sect.\,\ref{sec:rsg_observations}).
For constant overshooting, we conclude that the synthetic population with $\alpha_\mathrm{sc} \approx 100$ and $\alpha_\mathrm{ov} = 0.33$ performs best on simultaneously explaining the RSG and BSG populations in the SMC. In this case, a b/r ratio of the order of one is predicted (162 RSGs per 138 BSGs, see Figs.\,\ref{fig:l_dist} \& \ref{fig:l_dist_bsg}). Most other combinations of efficiencies predict either a very low or very high b/r ratio.

While these values might provide the most satisfying match between observations and theory, there could be a mismatch at luminosities below the range that we discussed above. The evolutionary sequences with $M_\mathrm{ini} = 12$\Msol and below do not experience blue loops for these values, while some A- and B-type supergiants are observed in the associated luminosity range. We note that for these relatively faint stars, detection biases are larger. A possible explanation is that the extent of overshooting is lower at lower initial masses, as proposed by \cite{Castro14} and Grin et al. (in prep. -- see also Sect.\,\ref{sec:ms_evolution}). If that is the case, stars with lower initial masses also become BSGs (Fig.\,\ref{fig:six_hrds} and \ref{fig:thirty_hrds}). Alternatively, these objects could have a binary origin -- we discuss this in Sect. \ref{sec:validity}. To be able to draw any strong conclusion, a more complete observational picture is warranted.

\subsection{Surface abundances \label{sec:n_surface}}
The nitrogen abundance at the surface of a star can be used as a probe for its evolutionary history. 
We illustrate this in Fig.\,\ref{fig:hrd_xns}. For stars that are approaching the red giant branch from hotter temperatures, 
no nitrogen enhancements are predicted by our evolutionary sequences in the non-rotating case (shown in the left panel). 
However, for stars that are on a blue loop excursion, it is predicted that CNO-processed material is dredged up 
from their deep convective envelopes during the prior RSG phase. As a result, the mass fractions of helium and nitrogen at the surface are enhanced. This applies to the light blue diamonds in Fig.\,\ref{fig:hrd_xns} (left) around $\log (L / L_\odot ) = 5$ and $\log ( T_\mathrm{eff} / K ) = 4.2$. 
The evolutionary tracks that we show are those where $\alpha_\mathrm{sc} = 100$ and $\alpha_\mathrm{ov} = 0.33$, which could best explain the RSG and BSG populations.

When rotation is taken into account, also stars that have not developed deep convective envelopes can be enriched in nitrogen. This is illustrated in the right panel of Fig.\,\ref{fig:hrd_xns}, which shows evolutionary models with an initial rotation velocity of 300\kms. Still, stars that are are on a blue loop excursion are more enriched in nitrogen because an additional mechanism is at work. These are the orange diamonds around $\log (L / L_\odot ) = 5$ and $\log ( T_\mathrm{eff} / K ) = 4.2$ in the right panel of Fig.\,\ref{fig:hrd_xns}.

In practice, using nitrogen abundances to constrain the evolutionary history of stars is difficult, 
because even on the MS a large fraction of stars show nitrogen enrichment 
\citep{Hunter08, Bouret13, Grin17}, and 
the observed nitrogen enrichment does not, or not clearly, correlate with rotation velocity. 
Yet, as shown in Fig.\,\ref{fig:hrd_xns}, also in the case with rotational mixing included, the BSGs which did previously undergo
an RSG dredge-up phase are significantly more enriched in nitrogen than those which did not.
A comprehensive view of the nitrogen abundances of the SMC supergiants would thus constitute a 
very constraining test of our models.


The nine A-type stars in the SMC analyzed by \cite{Venn99} do show nitrogen enrichment, varying from weak to strong. Because they all have luminosities below $\log(L / L_\odot ) \approx 5$, this is in agreement with our predictions shown in Fig.\,\ref{fig:hrd_xns} although less scatter would be expected.

\subsection{The most massive stars}

So far we have paid relatively little attention to the predictions of our models for the most luminous stars. The reason is that
such stars are expected to be quite rare, and there are very few solid observational constraints available
at this time. In addition, even though the SMC offers the advantage of mass loss rate uncertainties
being less important, this is not the case any more if we consider initial stellar masses of 60\Msol and higher.
Additionally, even in the SMC, such very massive stars are very close to the Eddington limit, such that envelope
inflation may strongly affect their radii and effective temperatures.  
Therefore, even when picking the model grid with the most promising mixing parameters, the models for
the highest masses are the least likely to be close to reality.

An exception is perhaps the core hydrogen burning phase of evolution, for which the mass loss rates are
the best understood. In this respect, our prediction that the TAMS bends to very cool temperatures 
at luminosities below $\log (L / L_\odot ) \approx 6$ for what we identified as the most likely mixing parameters
appears relatively robust (i.e., efficient semiconvection and intermediate overshooting). 
As shown in Fig.\,\ref{fig:six_hrds} and \ref{fig:thirty_hrds},
this leads to a continuous effective temperature distribution of the most luminous stars, 
rather than showing a pronounced post-MS gap as visible in the grids computed with the smallest overshooting parameters.
We note that \cite{Castro18} find tentative evidence for this; however the interpretation of their data is still somewhat ambiguous.

Finally, we want to point out that with the mass loss rate assumptions adopted for our models, which are state-of-the-art but are surely uncertain at low effective temperatures, there are essentially no hydrogen-poor or hot WR stars produced independent of our assumtions on internal mixing.

\section{Discussion \label{sec:validity}} 

\subsection{Summarizing our results \label{s}}

In the sections above, we have worked out the systematic trends that occur in massive star evolution models with an SMC initial chemical composition as function of the efficiency of internal mixing processes.
We have seen that these mixing processes determine the core mass increase and the slope of the H/He-gradient at the core-envelope interface, which are the internal structure parameters ruling the evolution of the stars in the HRD.

We have then confronted these evolutionary models with various observational constraints. We found that for the overshooting parameter $\alpha_\mathrm{ov}$ (which determines the increase of the core masses)
and the semiconvection parameter $\alpha_\mathrm{sc}$ (which is the key parameter for determining the H/He gradient) only a limited range of combinations is compatible with those observational constraints. 

We have seen that the constraints on the MS imply convective core overshooting 
of the order of $\alpha_\mathrm{ov}= 0.33$ 
in the considered mass range.
Interestingly, larger overshooting has the consequence that the stellar tracks produce essentially no BSGs with luminosities of $\log (L / L_\odot) \simle 5.5$ and temperatures $\log (T_{\rm eff}/K) \simle 4.3$ (cf., Fig.\,\ref{fig:thirty_hrds}). Since plenty of such stars are observed in the SMC \citep[e.g.,][]{Blaha89}, an overshooting parameter of $\alpha_\mathrm{ov}$ significantly above 0.33 seems to be excluded. 
This provides an independent constraint of the overshooting parameter, in addition to those derived from the width of the MS.
When the requirement of a large H/He-gradient from \cite{Schootemeijer18} is folded in, such large overshooting is ruled out even once more.
Similarly, $\alpha_\mathrm{ov} = 0\dots 0.11$ yields only BSGs for a wide range of initial masses for models where semiconvection is reasonably efficient. As a result, practically no RSGs are then predicted to exist in the luminosity range where \cite{Davies18} report 150 of these objects.

Conversely, Fig.\,\ref{fig:thirty_hrds} shows that inefficient semiconvection ($\alpha_\mathrm{sc} < 1$) can be
ruled out, because -- independent of the amount of overshooting -- far too few BSGs are produced. Again, this conclusion is reinforced by the fact that $\alpha_\mathrm{sc} < 1$ does not allow for steep H/He-gradients. In fact, the nearly complete overlap in the parameter
space of only shallow H/He-gradients (Fig.\,\ref{fig:dxdq_sup10}) with the parameter space where only RSGs are produced (Fig.\,\ref{fig:rsg_frac}) reinforces the conclusion of \cite{Schootemeijer18} that steep H/He-gradients are observationally required and generalizes it to a wider mass range. Finally, we have seen in Sect.\,\ref{sec:rsg_observations} that the parameter range that 
can at least partially explain the problem of the paucity of luminous RSGs falls within the range that is compatible
with all other observational constraints.

\begin{figure}
   \centering
   \includegraphics[width = \linewidth]{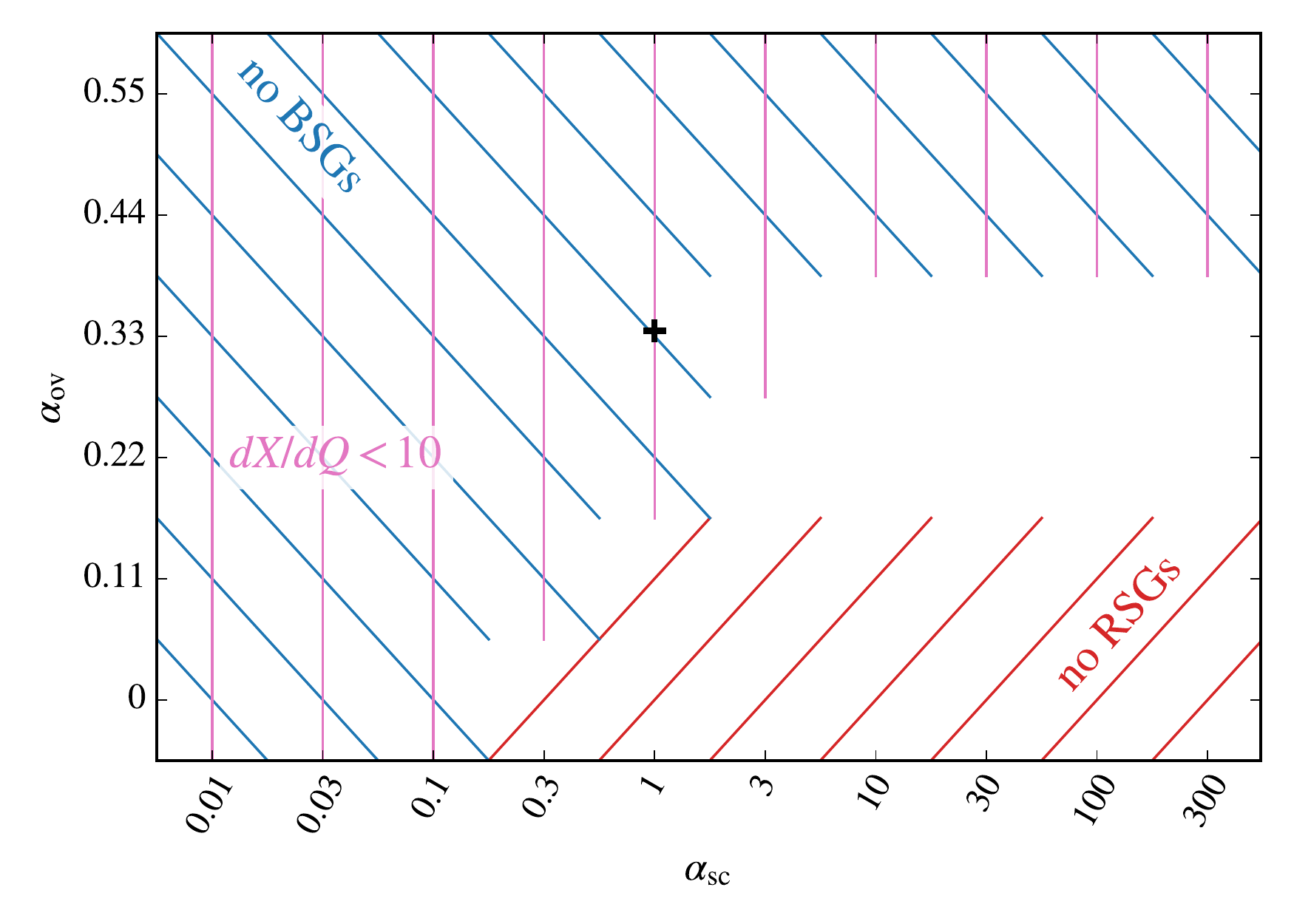}
   \caption{Schematic diagram indicating which part of the \asc and \aov parameter space is at odds with our 
three different criteria. These criteria are 1) the ability to produce steep H/He gradients (Sect.\,\ref{sec:hhe_grad}, 
purple vertical lines), 2) for stars with $\log ( L / L_\odot ) \lesssim 5.5$ to spend  
part of core helium burning as blue supergiant (blue diagonal lines), 
3) for stars with $\log ( L / L_\odot ) \geq 4.7$ during core helium burning to become red supergiants 
(red diagonal lines)
   .}
             \label{fig:overview}%
    \end{figure}

The situation is summarized in Fig.\,\ref{fig:overview}. We see that the parameter subspace where all observational tests are passed is rather small. In fact, the overshooting parameter seems to be well constrained to the interval $0.2 \simle \alpha_\mathrm{ov} \simle 0.35$. Semiconvection, on the other hand, is required to be efficient in the sense that $\alpha_\mathrm{sc} > 1$. Here it is to be mentioned that the allowed parameter space for the semiconvection parameter may appear larger than it is, since for  $\alpha_\mathrm{sc} \simgr 10$ our models show very similar behavior.

Importantly, Fig.\,\ref{fig:overview} shows
the existence of a subspace of the considered parameter space which appears to be compatible
with all observational constraints. We note that a priori there was no guarantee of this
outcome, which gives the hope that our problem is well posed -- in the sense that the adopted physical descriptions allow for a representation of the real stars -- despite the caveats discussed below. 
Further tests may show whether or not this turns out to be an illusion. For example, a study similar to this one needs to be done for Galactic and LMC composition, although the higher metallicity of these environments
may introduce additional uncertainties through the increased importance of stellar winds. The concept of this work may also be extended to intermediate mass stars. However, it is already clear that a consideration of a mass dependence of the mixing parameters can not be avoided in this case (see below).


\subsection{Caveats \label{c}}

Figure\,\ref{fig:overview} gives an overview of our findings, without subdividing the mass range 
we consider (9$\dots$100\Msol). However, in
the discussion above, we already had to specify smaller initial mass ranges 
when discussing specific observational constraints.
Indeed, the physical situation in our models changes significantly, with our 9\Msol models being
dominated by ideal gas pressure, while radiation pressure is more important in our 100\Msol models.

In fact, when considering the overshooting parameter, it already appears evident that (in the way it is currently parametrized) it can not be constant for all core hydrogen burning stars. For intermediate and low mass stars, good constraints exist 
pointing to values of $\alpha_\mathrm{ov} \simle 0.2$ \citep[e.g.,][]{Aerts13, Claret16}. 
On the other hand, the results of \cite{Castro14} argue for an increase of $\alpha_\mathrm{ov}$ with mass (Grin et al., in prep.). Furthermore, as discussed in Sect.\,\ref{sec:bsg_observations}, a lower overshooting than indicated by Fig.\,\ref{fig:overview} may yield better agreement with the blue-to-red supergiant ratio at the lowest considered masses.

Such a mass dependence would actually not be surprising, since 
$\alpha_\mathrm{ov}$ is an ad hoc parameter, which is not backed up by any physical theory.
The treatment of semiconvection, on the other hand, is based on a local, 
linear stability analysis \citep{Kato66, Langer83}, which fully
accounts for a mixture of gas and radiation pressure. Therefore, we may hope that the mass dependence of this parameter is weaker or absent.

Furthermore, our discussion of the chemical structure of massive stars remained limited, since we only considered the parameters core mass and H/He gradient. Clearly (as also indicated by Fig.\,\ref{fig:vrot_profs}) hydrogen and helium profiles can be quite complex and may need more than two parameters to describe them. Obviously, the inclusion of additional parameters would hardly be feasible at the moment. In any case, its necessity has not yet been shown, and the convergence of the viable part of the parameter space from multiple constraints is not arguing for it. 
  
On the other hand, we know that even for a fixed initial chemical composition,
the initial stellar mass is not the only parameter describing its future evolution. 
Rotation and binarity are widely discussed as initial parameters affecting the evolution of massive stars \citep{Maeder12, Langer12}. However, while for massive stars living in a binary is the rule rather than the exception \citep{Sana12}, the fraction of isolated massive stars whose evolution is 
significantly affected by rotation is unclear. As discussed in Sect.\,3, we find the effects
of rotation on our models to be quite limited -- except for extreme rotation, which allows for chemically homogeneous evolution \citep{Yoon06, Brott11}, but this is thought to be very rare. However, in the framework of models which allow for a significant redistribution of hydrogen and helium for average rotation rates (cf., Sect.\,3), rotation would need to be considered as an important third parameter.

Our neglect of binarity, however, is harder to justify, except for feasibility reasons. The two additional,
necessary initial parameters (initial secondary mass, and initial separation) tremendously blow up the initial parameter space
to consider. However, now that our analysis of single star models has narrowed the viable parameter space for the mixing parameters, it will be our next step to compute binary evolution grids with the current best guess for these, and investigate whether or not 
binarity affects our conclusions. 

While binarity may be omnipresent in massive stars, this does not imply per se that it is clearly important
for our discussion. Considering MS stars, of the order of 10\% and 15\% may be merger products or mass gainers in post mass transfer systems, respectively \citep{deMink14}. However, if these would be fully rejuvenated, they might be rather indistinguishable from ordinary single stars and evolve further on as such.
In fact, the very efficient semiconvective mixing advocated by our results would imply that rejuvenation occurs in the vast majority of cases (\cite{Braun95}). On the other hand, it has been suggested that stellar mergers produce strong magnetic fields \citep{Ferrario09, Langer12, Schneider16}, 
in which case the evolution on the MS and beyond could be strongly affected \citep{Petermann15}.
As about 7\% of the massive MS stars are found to be magnetic \citep{Fossati15, Grunhut17}, this may affect our analysis at this level.

The only other branch of massive binary evolution which may be of relevance here (because, e.g., common envelope evolution or
binaries involving compact objects produce exotic, easily identifiable types of stars) are post-MS stellar mergers. Such mergers produce stars whose core mass is smaller compared to that of a single star of the same mass. Such objects are known to spend nearly all of core helium burning as BSGs \citep{Braun95, Justham14}. The fraction of massive binaries that produces such stars is again estimated to be of the order of 10\% \citep{Podsiadlowski92}; however, the uncertainty of this number is considerable \citep{deMink14} and not independent of the mixing parameters that we discuss here.

In summary, there is much more work to do to consolidate or modify our conclusion. 
However, at present it is reassuring that many constraints seem to point in the same direction.


\section{Conclusions \label{sec:conclusions}}
In this study, we have computed a large number of grids of massive star model sequences to comprehensively analyze 
the effects of the most relevant internal mixing processes on their hydrogen burning and post main sequence evolution. We chose to focus on models with an initial chemical composition of that
of the Small Magellanic Cloud (SMC), such that uncertain mass loss rates affect our conclusions as little as possible.

We compared the predictions of our models to a multitude of observational constraints in the SMC. These included the observed widths of the main sequence band in the Hertzsprung-Russell diagram, the presence of both blue and red supergiants for a considerable range of luminosities, the empirical upper luminosity limit of red supergiants, and the requirement of
a steep H/He-gradient at the core-envelope interface in the hot hydrogen-rich Wolf-Rayet stars. 

As summarized in Sect.\ref{s} and Fig.\,12, we find a small
and well-defined subspace of the mixing parameters where all constraints can be satisfied simultaneously.
Conversely, we can exclude most of the considered parameter space. In terms of the formulation of the
mixing processes used in our models, we find that semiconvective mixing needs to be efficient
($\alpha_\mathrm{sc} \simgr 1$), while convective core overshooting is restricted to intermediate
values ($0.22 \simle \alpha_\mathrm{ov} \simle 0.33$) for most of the considered mass range, 
with possibly smaller values favoured below $\sim 10\mso$.
Rotational mixing in our models was found of limited importance.

In terms of structural parameters, which are relevant beyond our chosen mathematical formulation of the mixing processes, 
we find a necessity of a moderate core mass increase over the canonical value, 
which may be produced by convective core overshooting or otherwise, and a H/He-gradient 
which is at least five times steeper than the one that emerges naturally from the retreating 
convective core during core hydrogen burning. While the latter has been concluded 
previously for stars above $\sim$40\Msol, we find this here to be valid stars down to $\sim$9\Msol.

Since our results could only be derived with various caveats (cf. Sect.\ref{c}), more theoretical work is
needed to consolidate them, in particular to include models of close binary evolution. In any case,
much more stringent tests of our results will be enabled by a complete set of spectroscopic data 
of the massive stars in this key galaxy, the SMC, which is currently available only fragmentary.




\bibliographystyle{aa} 
\setlength{\bibsep}{0pt}
\bibliography{bib.bib}{}

\clearpage

\onecolumn

\begin{appendix}
\section{Extra Hertzsprung-Russell diagrams\label{app:a}}


   \begin{figure*}[ht]
   \centering
   \includegraphics[width = 0.9\linewidth]{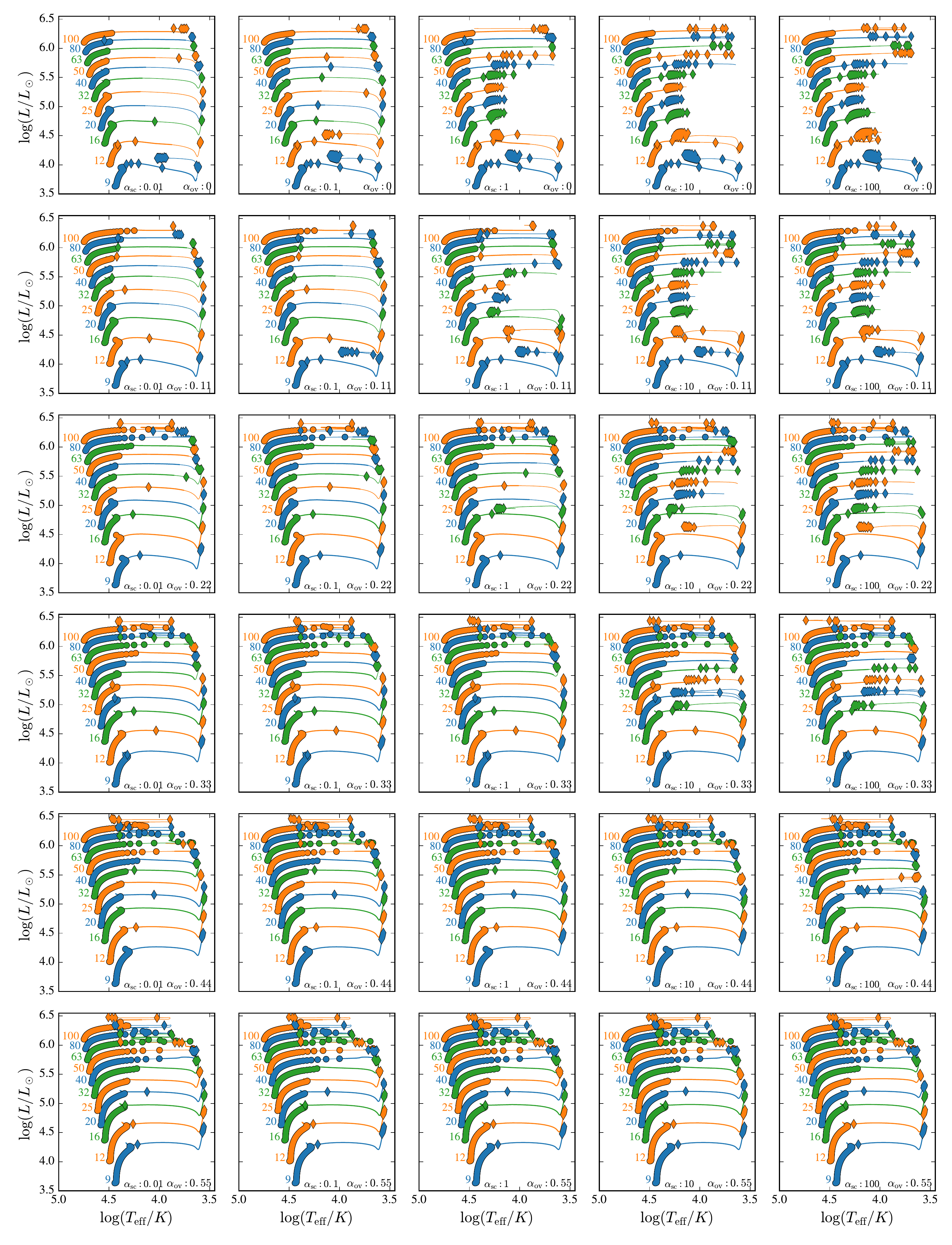}
   \caption{Same as Fig.\,\ref{fig:six_hrds}, but now we show more combinations of \asc and \aov.}
             \label{fig:thirty_hrds}%
    \end{figure*}


   \begin{figure*}[ht]
   \centering
   \includegraphics[width = 0.9\linewidth]{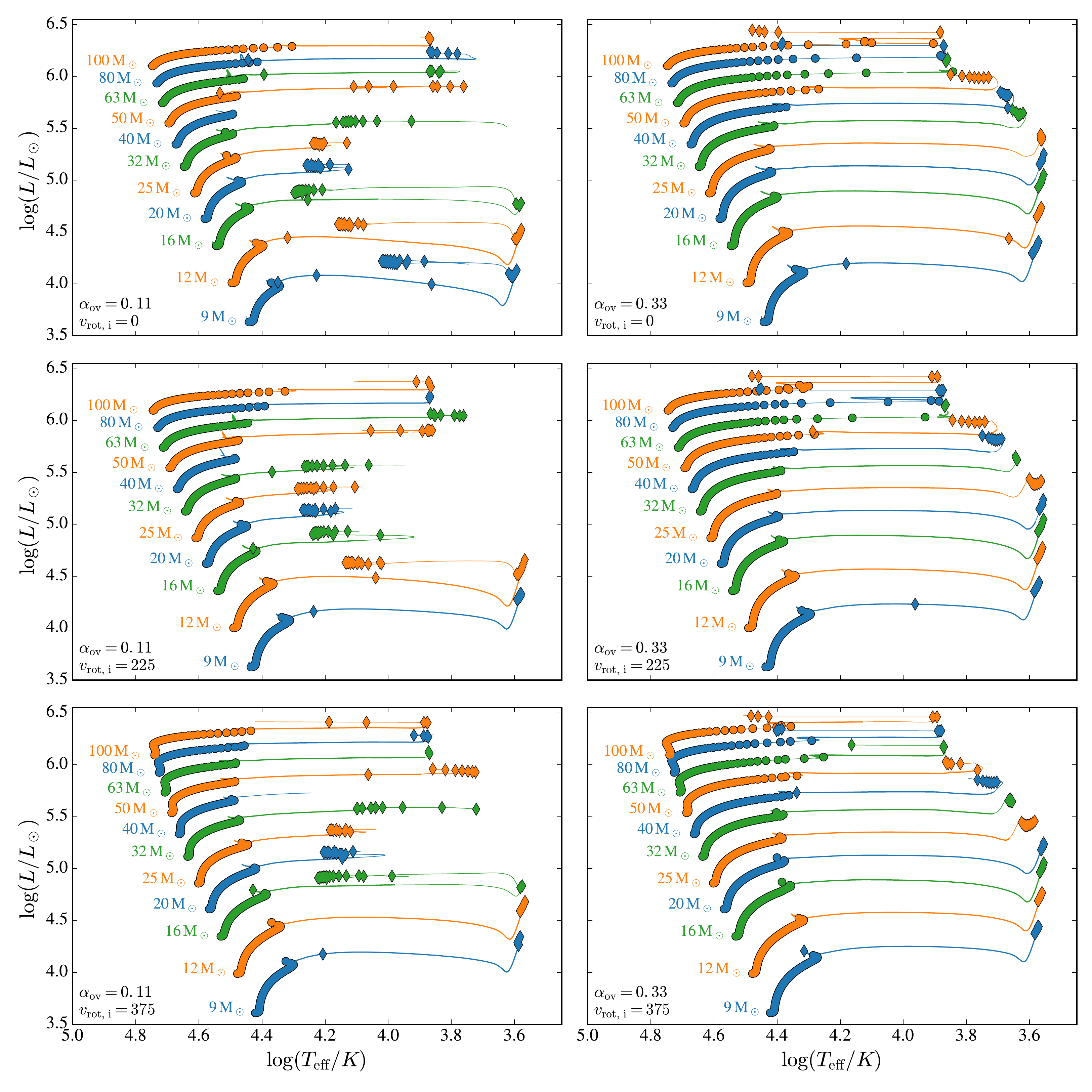}
   \caption[width = \linewidth]{Same as Fig.\,\ref{fig:six_hrds}, but now we vary \aov and the initial rotation velocity (indicated in the bottom left of each panel, in units of \kms ). The models were computed with a \textit{relatively} high time resolution (see Sect.\,\ref{sec:method}) and with $\alpha_{\rm sc}=1$.}
             \label{fig:six_hrds_rot}%
    \end{figure*}

\end{appendix}

\end{document}